\documentclass[prd,showkeys,preprintnumbers,floatfix,notitlepage,
nofootinbib,superscriptaddress]{revtex4-1}
\usepackage{float}
\usepackage{slashed}
\usepackage{mathtools}
\usepackage{amsfonts} % AMS
\usepackage{amssymb} % AMS
\usepackage{amsmath} % AMS
\usepackage{array} % array
\usepackage{dcolumn} % Align table columns on decimal point
\usepackage{bm} % bold math
\usepackage{latexsym} % latex symbols
\usepackage{longtable} % long tables
\usepackage{hyperref} % hypertext links 
\usepackage{verbatim}
\usepackage{epsfig}
\usepackage{color}

\usepackage[utf8]{inputenc} %% required by INSPIRE
\newcommand{\cI}[0]{\mathcal I}
\newcommand{\cJ}[0]{\mathcal J}
\newcommand{\cO}[0]{\mathcal O}
\newcommand{\cM}[0]{\mathcal M}

\newcommand{\cK}[0]{\mathcal K}
\newcommand{\cD}[0]{\mathcal D}
\newcommand{\cS}[0]{\mathcal S}
\newcommand{\cC}[0]{\mathcal C}
\newcommand{\cL}[0]{\mathcal L}
\newcommand{\cR}[0]{\mathcal R}

\newcommand{\tr}[0]{{\rm tr}}
\newcommand{\wt}[0]{\widetilde}
\newcommand{\PV}[0]{{\rm PV}}
\newcommand{\iso}[0]{{\rm iso}}
\newcommand{\df}[0]{{\rm df}}
\newcommand{\thr}[0]{{\rm thr}}
\newcommand{\Kiso}[0]{{\cK_{\df,3}^{\iso}}}
\newcommand{\Mthr}[0]{{\cM_{3,\thr}}}
\newcommand{\Mdf}[0]{{\cM_{\df,3}}}
\newcommand{\Mdfthr}[0]{{\cM_{\df,3,\thr}}}
\newcommand{\Kdf}[0]{{\cK_{\df,3}}}
\newcommand{\Fiso}[0]{{F_{3}^{\iso}}}

\newcommand{\SPT}[0]{Sharpe:2017jej}
\newcommand{\HSBS}[0]{Hansen:2016ync}
\newcommand{\HSPT}[0]{Hansen:2015zta}
\newcommand{\HSTH}[0]{Hansen:2016fzj}
\newcommand{\BHSQC}[0]{Briceno:2017tce}
\newcommand{\HSQCb}[0]{Hansen:2015zga}
\newcommand{\HSQCa}[0]{Hansen:2014eka}

\newcommand{\Akakia}[0]{Hammer:2017uqm}
\newcommand{\Akakib}[0]{Hammer:2017kms}
\newcommand{\Luscher}[0]{Luscher:1986n2,Luscher:1991n1}
\newcommand{\AkakiBS}[0]{Meissner:2014dea}

\begin{document}
 \preprint{\vbox{\hbox{JLAB-THY-18-2657} }}
  \preprint{\vbox{\hbox{CERN-TH-2018-046} }}
\title{Numerical study of the relativistic three-body quantization condition\\ in the isotropic approximation}
%

%%%%%%%%%%
\author{Ra\'ul A. Brice\~no}
\email[e-mail: ]{rbriceno@jlab.org}
\affiliation{Thomas Jefferson National Accelerator Facility, 12000 Jefferson Avenue, Newport News, VA 23606, USA}
\affiliation{ Department of Physics, Old Dominion University, Norfolk, VA 23529, USA}
%%%%%%%%%%

%%%%%%%%%%
\author{Maxwell T. Hansen}
\email[e-mail: ]{maxwell.hansen@cern.ch}
\affiliation{Theoretical Physics Department, CERN, 1211 Geneva 23, Switzerland}%CERN-TH, CH-1211, Geneva, Switzerland}
%%%%%%%%%%

%%%%%%%%%%
\author{Stephen R. Sharpe}
\email[e-mail: ]{srsharpe@uw.edu}
\affiliation{Physics Department, University of Washington, Seattle, WA 98195-1560, USA\\}
%%%%%%%%%%

\date{\today}

\begin{abstract}
We present numerical results showing how our recently proposed relativistic three-particle
quantization condition can be used in practice. Using the isotropic (generalized $s$-wave)
approximation, and keeping only the leading terms in the effective range expansion,
we show how the quantization condition can be solved numerically in a straightforward
manner.
In addition, we show how the integral equations that relate the intermediate
three-particle infinite-volume scattering quantity, $\Kdf$, to the physical scattering
amplitude can be solved at and below threshold.
We test our methods by reproducing known analytic results for the $1/L$ expansion of the threshold state,
the volume dependence of three-particle bound-state energies,
and the Bethe-Salpeter wavefunctions for these bound states. 
We also find that certain values of $\Kdf$ lead to unphysical finite-volume energies, 
and give a preliminary analysis of these artifacts. 
%Future work is required to fully understand their origin.
\end{abstract}

\maketitle

\section{Introduction}

Studies of few-hadron systems based on lattice quantum chromodynamics (LQCD) are advancing rapidly. Recent results highlighting this progress include the first study of multiple, strongly-coupled scattering channels~\cite{Dudek:2014qha, Woss:2018irj}, the first determination of resonant electroweak amplitudes~\cite{Feng:2014gba,Briceno:2016kkp}, and the first study of a meson-baryon scattering amplitude in a resonant channel~\cite{Andersen:2017una}. %
Each of these calculations has been made possible by a series of theoretical developments, stemming from seminal work by L\"uscher~\cite{\Luscher}. This formalism and its subsequent generalizations explain how the desired infinite-volume observables, namely scattering and transition amplitudes, can be obtained from the finite-volume correlation functions evaluated using numerical LQCD. We point the reader to Ref.~\cite{Briceno:2017max} for a recent review on the topic. 

Current theoretical work is focused on extending the finite-volume relations to
extract observables with initial or final states composed of three or more hadrons. To this end, in a series of papers published in the last few years, we have derived  a quantization condition that relates the
finite-volume energies of states containing a three-particle component to infinite-volume, two- and three-particle 
scattering amplitudes~\cite{\HSQCa,\HSQCb,\BHSQC}.%
\footnote{For parallel studies of three-body systems see Refs.~\cite{Kreuzer:2008bi,Kreuzer:2012sr,Briceno:2012rv, Polejaeva:2012ut,Guo:2016fgl,\Akakia,\Akakib, Mai:2017bge, Guo:2017ism, Doring:2018xxx}. }
This quantization condition accounts for all power-law volume dependence while dropping dependence that falls exponentially with the box length, $L$.
The formalism is relativistic and encompasses arbitrary interactions aside from two
restrictions:
(i) the particles must be spinless and identical, and 
(ii) the two-particle K matrix cannot have poles in the kinematical regime of interest.
From our past experience in the two-body sector~\cite{Briceno:2014oea, Briceno:2015csa}, we expect the former restriction to be straightforward to remove, and now understand
how to remove the latter~\cite{BHS3}.
The relation to physical scattering amplitudes involves two steps.
In the first, the quantization condition is used to determine an infinite-volume
K matrix like quantity, $\Kdf$~\cite{\HSQCb}.
In the second, $\Kdf$
is related to the physical scattering amplitudes via integral equations.\footnote{%
We stress that these integral equations are defined via manifestly finite integrals with fixed total three-particle energy. In addition, the equations depend only on on-shell quantities and make no reference to an underlying effective theory.}
$\!\!\!\!^,$\footnote{
In general, taking these steps will require using parametrizations for the physical scattering amplitudes, such as those currently being developed in Ref.~\cite{Mai:2017vot}.}
The formalism has been tested in several ways, most notably by reproducing the known finite-volume dependence of a weakly-interacting threshold state and of an Efimov-like bound state~\cite{\HSTH,\HSPT,\SPT,\HSBS}.

A crucial issue yet to be considered, however, is whether the formalism is usable in practice.
Indeed, in recent papers introducing an alternative approach based on 
nonrelativistic effective field theory (NREFT),
Refs.~\cite{\Akakia,\Akakib} have suggested that our formalism may be too complicated
to use in the analysis of real lattice data. 
It is the purpose of this work to investigate this issue.
We find, in fact, that the status with regard to applicability is more-or-less identical
to that for the NREFT approach: the steps are the same, the number of parameters are the same
(when using analogous approximations),
and the numerical implementation seems to be of comparable difficulty.
There are, however, technical differences that we 
discuss briefly here and return to in the conclusions.\footnote{%
The steps in our approach are also similar to those in the recent
relativistic proposal of Ref.~\cite{Mai:2017bge}.
This parametrizes three-particle interactions using an isobar (dimer) formalism that
maintains unitarity. This parametrization is then used both in finite volume to
predict the spectrum, and, in a separate calculation, in infinite volume to give the
scattering amplitude. We suspect that this formalism will yield similar results to ours.}

In particular, we note here four advantages of using a relativistic formalism for
three-particle physics.
First, one aims to constrain the physical observables over the widest energy range 
possible, and our formalism applies for three-particle center-of-momentum (c.m.)
energies reaching up to 
$4m$ ($5m$ if there is a $\mathbb Z_2$ symmetry forbidding odd-legged vertices), 
clearly in the regime of relativistic momenta. Second, in Ref.~\cite{\BHSQC} we describe how to 
determine $\textbf 2 \to \textbf 3$ scattering amplitudes from finite-volume energies. 
Such processes are  intrinsically relativistic since the incoming particles must have 
enough kinetic energy to produce a new  particle. 
Third, it is known in the $1/L$ threshold expansion that, for weakly interacting systems, 
three-body effects and relativistic effects enter at the same order in $1/L$. Thus it is natural to 
pursue a formalism that includes both. Fourth, as we describe below, for three noninteracting 
particles the second and third excited states (as well many higher groups of states) become 
degenerate in the nonrelativistic limit. Thus the basic counting and locations of noninteracting 
states, as well as their deformations due to interactions, is very different between the relativistic 
and nonrelativistic theories. This final point is discussed further
in Sec.~\ref{sec:Kisozero}.

In this work, to address the issue of applicability, we primarily use a dynamical approximation similar to that used
in the numerical example worked out in Ref.~\cite{\Akakib}, referred to here as
the low-energy isotropic approximation. However, in all calculations presented here, we make no kinematical approximations, i.e.~we keep the relativistic form throughout. In addition, we restrict attention to theories in which there is a $\mathbb Z_2$ symmetry
forbidding transitions between even- and odd-particle-number sectors. This is a simplifying approximation that we know, at least formally, how to remove~\cite{\BHSQC}. In short, we conclude that the three-body formalism we have previously derived~\cite{\HSQCa,\HSQCb,\BHSQC} is indeed in a form that is suitable for the analysis of some realistic lattice systems. 

The remainder of this paper is organized as follows. 
In Sec.~\ref{sec:formalism} we present a brief summary of the three-body formalism,
and explain the justification for the isotropic approximation, in which
the matrix quantity, $\Kdf$, is replaced by a single function of the total three-particle energy, $\Kiso$.
In Sec.~\ref{sec:results} we present several results concerning the three-particle
spectrum obtained using the quantization condition, starting with the simplest case
of vanishing $\Kiso$ and then turning on nonzero values. 
In cases where this leads to a three-particle bound state,
we compare the volume dependence of the bound-state energy to an analytic prediction. 
We close Sec.~\ref{sec:results} by
studying the volume dependence of the threshold state and comparing it to
analytic predictions.
In Sec.~\ref{sec:thresh} we implement the relation between $\Kiso$ and the physical scattering 
amplitude, beginning below threshold and then working directly at threshold.
This illustrates how our complete, two-stage formalism can be implemented.
In Sec.~\ref{sec:unphys} we describe how, in certain regimes of parameters, unphysical
solutions to the quantization condition can appear, and we discuss their possible origin.
We conclude and describe directions for future work in Sec.~\ref{sec:conclusion}.
Two appendices describe some technical details of our numerical implementation of
the quantization condition and our methods for solving the integral equations.

\section{Summary of formalism in the isotropic low-energy approximation \label{sec:formalism}}

In this section we recall the essential results for the $\mathbb Z_2$-symmetric case; 
further details can be found in Refs.~\cite{\HSQCa,\HSQCb}.
The spatial volume is a cube of length $L$ with periodic boundary conditions,
so that finite-volume momenta have the form $\vec k=2\pi \vec n/L$, 
with $\vec n$ a three-vector of integers.
The total momentum, $\vec P$, can take any value in this finite-volume set.

Within this set-up, the result of Ref.~\cite{\HSQCa} is that, 
for any fixed values of $L$ and  $\vec P$, 
the finite-volume energy spectrum, $\{E_n(L)\}$, 
is given by solutions to the quantization condition\footnote{%
The ultimate aim is for this result to be used to interpret results from lattice
QCD simulations. These results inevitably involve errors due to working at nonvanishing
lattice spacing. Such effects are not incorporated into the quantization condition, which is a
continuum quantum field theory result. Thus, strictly speaking,
lattice results should be extrapolated to the continuum limit before they can be
used in the quantization condition.}
\begin{equation}
\det \big [F_3(E, \vec P, L) + \cK_{\df,3}^{-1}(E^*) \big ] = 0\,.
\label{eq:QC}
\end{equation}
Here the finite-volume-frame energy, $E$, is related to the c.m.-frame 
energy, $E^*$, by the standard dispersion relation, $E^{*2} = E^2 - \vec P^2$. 

In Eq.~(\ref{eq:QC})
the quantities $F_3$ and $\Kdf$ are matrices in a space 
labeled by the finite-volume momentum, $\vec k$, of one of the particles
(denoted the ``spectator")
and the angular momentum of the
other two in their two-particle c.m.~frame. The determinant above acts on this space.
$\Kdf$ is an infinite-volume quantity characterizing the 
underlying local three-particle interaction. It is analogous to the three-body contact terms in the NREFT approach of Ref.~\cite{\Akakib}. 
$F_3$ incorporates both the effects of two-particle scattering and of the finite volume.
More specifically, it depends on the two-particle K matrix, $\cK_2$, 
and on known kinematic finite-volume  functions. 
Its explicit form is given in Eq.~(\ref{eq:F3s}) below in the approximation we use.

Just as in the two-body sector~\cite{\Luscher, Kim:2005gf, Hansen:2012tf, Briceno:2012yi}, 
to use the quantization condition 
in practice one must truncate the partial waves that contribute, 
thus reducing the matrices to finite size~\cite{\HSQCa}. 
This applies here not only to $\cK_2$, as in the two-particle case, but also to $\Kdf$.
One then proceeds as follows:\footnote{%
This description applies to theories with a $\mathbb Z_2$ symmetry. For the general case there
are more quantities to determine but the overall approach is the same~\cite{\BHSQC}.}
\begin{enumerate}
\item Perform a two-particle
finite-volume analysis to determine $\cK_2$ as a function of the two-particle
c.m.~energy, $E^*_2$, using L\"uscher's quantization condition~\cite{\Luscher} and its 
generalizations.
\item
Use the quantization condition, Eq.~(\ref{eq:QC}), and the three-particle spectrum to 
constrain $\Kdf$.
\item
Determine the relativistic three-to-three scattering amplitude, $\cM_3$, 
from $\Kdf$ and $\cK_2$ by solving the integral equations given in Ref.~\cite{\HSQCb}. 
\end{enumerate}
Our aim here is to show how this procedure works when we truncate to 
a single partial wave and make a few further simplifying approximations.

An important technical point is that our formalism includes
a smooth cutoff function, $H(\vec k)$, that depends on the spectator momentum $\vec k$.
For fixed $E$, as $\vec k$ is increased the c.m.~energy in the remaining two-particle subsystem,
$E_{2,k}^*$, decreases, dropping first below the two-particle threshold and eventually 
becoming complex. 
Our formalism requires that $E_{2,k}^*$ is real and positive, $E_{2,k}^*> 0$, and the cutoff
function ensures that this condition is satisfied. 
This means that the sum over $\vec k$ is truncated to a finite number of terms.

There are two reasons for requiring
$E_{2,k}^*  > 0$. First, $\cK_2$ has a singularity (the left-hand cut) at this point, and this
can lead to additional power-law finite-volume effects that are not accounted for in the
formalism. Second, the boost to the two-particle c.m.~frame becomes unphysical
if the condition is not satisfied. There remains, however, considerable latitude in the
choice of cutoff function. In particular, the lower limit on $E_{2,k}^*$  can lie anywhere
in the range from $0$ to $(2-\delta)m$, with $\delta$ a positive constant of order one.
The final results for physical quantities should be independent of this cutoff 
(up to terms suppressed by $e^{-\delta m L}$).
We stress that, if $\delta$ is order one, then the cutoff occurs for spectator momenta
satisfying $|\vec k| \sim m$ and thus lying in the relativistic regime.\footnote{%
In the NREFT approach of Refs.~\cite{\Akakia,\Akakib} there 
is no corresponding constraint on the sum over spectator momentum, 
nor is there a need for the cutoff to be smooth.
While this simplifies practical calculations, it comes at the price that physical
singularities such as the left-hand cut have to be dealt with in some fashion.} 
In this work we set $\delta=2$ throughout. 

\subsection{Definition and motivation of the isotropic approximation}

The approximation we consider here consists of three parts. First, we restrict $\cK_2$
and $\Kdf$ to contain only s-wave interactions between the nonspectator pair.
This implies that all matrices appearing in the quantization condition
have only the spectator-momentum indices, 
e.g.~$\cK_{\df,3} %=\cK_{\df,3;kp} 
= \Kdf(E^*, \vec k, \vec p\,)$. 
As noted above, these indices are truncated by the cutoff function.
Second, we assume that $\Kdf$ depends only on $E^*$ and not on the spectator momenta,
so that $\Kdf(E^*, \vec k, \vec p\,) \equiv \Kiso(E^*)$, independent of $\vec k$ and $\vec p$.
Together these give the ``isotropic approximation" introduced in  Ref.~\cite{\HSQCa}. 
Finally, we neglect the energy dependence of 
$q^*_2 \cot \delta(q^*_2)$ appearing within $\cK_2$.
This corresponds to taking only the leading order (scattering-length-dependent) 
term in the effective range expansion.

In the remainder of this section we explain why the isotropic approximation is the natural
generalization of the s-wave approximation in the two-body case. 
We begin by recalling the argument for the latter case.
We make use of the two independent Mandelstam variables, which
we denote by $s_2=4 q_2^{*\; 2} + 4 m^2$ and $t_2=-2 q_2^{*\;2} (1-\cos\theta)$, 
where $q^*_2$ is the magnitude of the c.m.~frame momentum.
The key input is that, at fixed $s_2$ and away from isolated poles, $\cK_2$ is a finite and thus square-integrable function of $\cos \theta$. This means that it admits a convergent decomposition in the Legendre polynomials, $P_\ell(\cos \theta)$. Alternatively, at fixed $s_2$, $\cK_2$ is an analytic function of $t_2$ near threshold so that one can perform a Taylor expansion about $t_2=0$. Combining these two expansions, we deduce that the coefficient of the $\ell$th polynomial, call it $\cK_{2,\ell}(q^*_2)$, must scale as $q^{*\; 2 \ell}_2$ as $q^*_2 \to 0$. This holds because the $\ell$th polynomial contains a term proportional to $\cos^\ell \theta$ and this must correspond to the $(t_2)^{\ell}$ term in the Taylor expansion.
Thus the s-wave contribution dominates close to threshold.

To justify the isotropic approximation in the three-body case, 
it is convenient to work with the full divergence-free K matrix, 
without the decomposition into interacting-pair partial waves.
This quantity is function of generalized Mandelstam variables, which we label
\begin{align}
s &= (p_1+p_2+p_3)^2\,, \\
s_{ij} &= (p_i+p_j)^2  \ \ {\rm and} \ \  s'_{ij} = (p'_i+p'_j)^2  \quad [i < j] \,,\\
t_{ij} &= (p_i-p'_j)^2\,,
\end{align}
where $i,j=1-3$, while $p_i$ are the initial and $p'_j$ the final four-momenta.
Note that, at threshold, $s=9m^2$, $s_{ij}=4m^2=s'_{ij}$, and $t_{ij}=0$.
There are many relations between these variables, so that, in addition to $s$, 
there are only seven independent kinematic variables.\footnote{%
One choice is $s_{12}$, $s_{13}$, $s'_{12}$, $s'_{13}$,
$t_{11},$ $t_{22}$, and $t_{33}$.}
For fixed $s$, the remaining variables are all ``angular", in the sense that they
span a compact seven-dimensional space~\cite{Weinbergbook}.
In particular, for fixed $s=9m^2+ \Delta$,  the quantities that measure the
distance from threshold, namely $\delta_{ij}\equiv s_{ij}-4m^2$, 
$\delta'_{ij} \equiv s'_{ij}-4m^2$ and $t_{ij}$, 
are all bounded in magnitude by $c \Delta$, where $c=\mathcal O(1)$.
This follows because of the relations
\begin{equation}
\sum_{i<j} \delta_{ij} = \sum_{i<j} \delta'_{ij} = -\tfrac12 \sum_{i,j} t_{ij} = \Delta\,,
\end{equation}
together with the fact that $\delta_{ij}$, $\delta'_{ij}$ and $-t_{ij}$ are all positive.

The key input now is that, at fixed $s$, $\Kdf$ should be an analytic function 
of the kinematic variables in the vicinity of the threshold. 
Performing a Taylor expansion about threshold, the leading term is independent of
$\delta_{ij}$, $\delta'_{ij}$ and $t_{ij}$, with the leading dependence on these variables proportional to $\Delta$.
Thus, close to threshold, the dominant contribution is independent of the angular variables.
One choice of these variables is given by those introduced in Ref.~\cite{\HSQCa},
namely the initial and final spectator momenta introduced above, $\vec k$ and $\vec p$,
together with the initial and final directions of the nonspectator pairs in their respective c.m.~frames,
$\hat a^*$ and $\hat a'^*$. These ten variables are reduced to seven 
by overall rotation invariance. 
Thus we conclude that the dominant near-threshold contribution is
not only independent of $\hat a^*$ and $\hat a'^*$ (which is the s-wave approximation for
$\Kdf$ already introduced above), but also of $\vec k$ and $\vec p$,
yielding the isotropic approximation.\footnote{%
It would be interesting to extend this argument to determine the form of the
$\mathcal O(\Delta)$ corrections in terms of $\vec k$, $\vec p$, $\hat a^*$ and $\hat a'^*$,
but this is beyond the scope of the present work.}

We close by commenting that, in the two-particle sector, the s-wave
approximation holds both for the K matrix, $\cK_2$, and the scattering-amplitude, $\cM_2$. 
Indeed the harmonic components of these two-objects have the same low-momentum scaling, 
the usual $(q^*_2)^{2\ell }$. This differs from the situation in the three-particle sector, 
where the argument holds for $\Kdf$ but fails for the scattering amplitude, $\mathcal M_3$.
The reason is that the latter exhibits kinematic singularities, 
discussed at length in Refs.~\cite{\HSQCa,\HSQCb}. 
In particular, $\cM_3$ is not smooth (indeed it diverges) at threshold 
and one cannot expect its harmonic coefficients to show low-energy suppression. 
This is a key advantage of $\Kdf$ over $\cM_3$.

\subsection{Quantization condition in the isotropic approximation}

\begin{figure}
\begin{center}
\includegraphics[width=\textwidth]{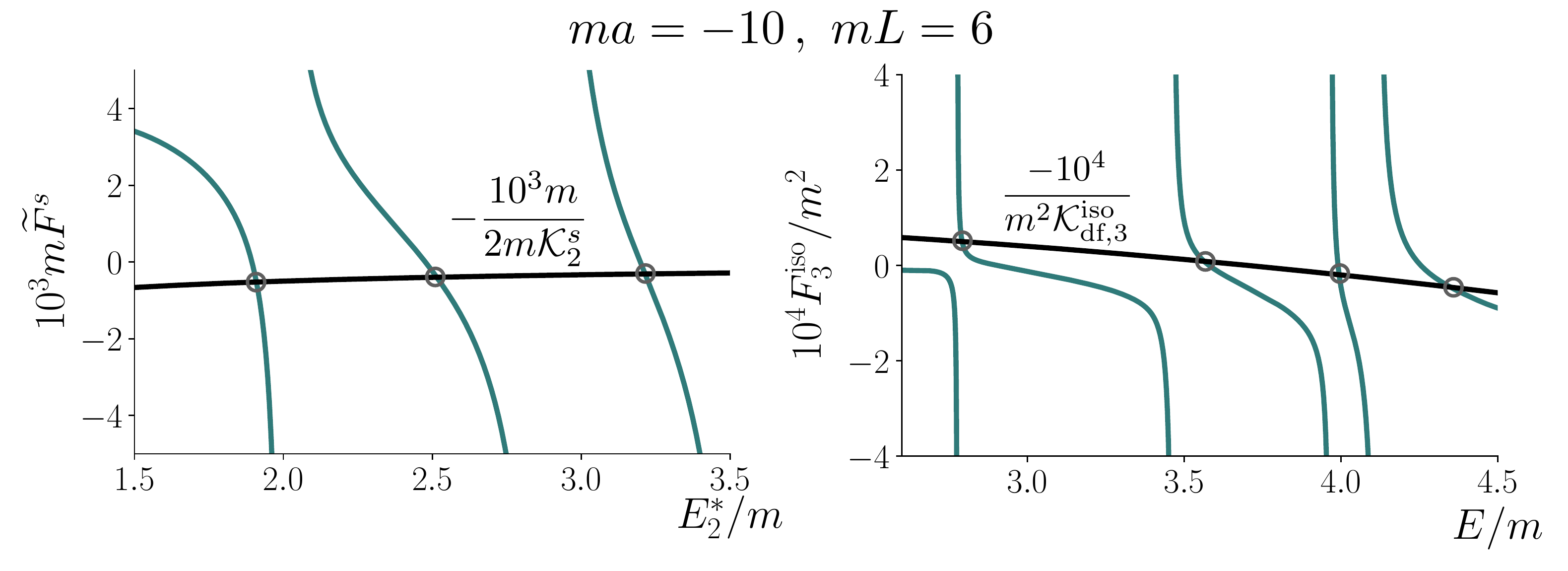}
\caption{Examples of solving the quantization conditions in the two-particle (left) and three-particle (right) sectors for $\vec P=0$ and $mL=6$. The two-particle condition in the left panel can be written as $\widetilde F_2^s = - 1/(2 \omega \mathcal K_2^s)$ where $\omega$ is the energy of the spectating third particle. This is satisfied when the two curves intersect, as indicated by the gray circles. (Here we take the spectator to have $\vec k=0$ and thus $\omega = m$.) This is closely analogous to the isotropic three-particle quantization condition given by Eq.~(\ref{eq:QCiso}), again satisfied at the indicated intersection points in the right panel. For this example, we take $\mathcal K_2^s$ from the leading order effective-range expansion with $ma=-10$, corresponding to an attractive two-particle interaction that pulls the lowest level below $E_2^*/m=2$. We take $1/\mathcal K_{\text{df},3}^\text{iso}$ to be a simple polynomial in $E/m$. 
\label{fig:QCexample}
}
\end{center}
\end{figure}

We now return to the main argument. As shown in Ref.~\cite{\HSQCa},
the isotropic approximation reduces the quantization condition to an algebraic equation
\begin{equation}
F_3^\iso(E, \vec P, L) = - 1/\Kiso(E^*)
\,.
\label{eq:QCiso}
\end{equation}
 To reach this form we first note that the determinant over angular momentum appearing in Eq.~(\ref{eq:QC}) is trivial given that only the $\ell=0$ contribution to the K matrix is nonzero. Second, in the isotropic approximation, the K matrix is independent of the spectator momentum. Therefore, the only eigenvector of $\Kdf$ in the space of spectator momenta with nonzero eigenvalue is that in which every entry is unity, i.e.~$|\textbf{1}\rangle=(1,1,\ldots,1)$.\footnote{%
The other eigenvectors, which have vanishing eigenvalues of $\Kdf$, lead to free three-particle
states, as discussed in Ref.~\cite{\HSQCa}.}
In this way only a one dimensional block of the matrices contributes, leading to
Eq.~(\ref{eq:QCiso}).
As noted above, this form is analogous to the s-wave approximation of the two-particle formalism.
In Fig.~\ref{fig:QCexample} we give an example of how this condition is used and compare 
to the s-wave two-particle case.

For any fixed $L, \vec P$ and any given finite-volume energy, $E_n(L, \vec P)$, Eq.~(\ref{eq:QCiso}) directly gives the value of $\Kiso(E^*)$ at $E^* = [{E_n(L, \vec P)^2 - \vec P^2}]^{1/2}$. This assumes that $\cK^s_2(E^*_2)$ is known for all $E_2^* < E^* - m$, as this is needed to determine $\Fiso$, defined below. Given $\Kiso(E^*)$, one can determine the corresponding $\mathcal M_3(E^*, \Omega_3', \Omega_3)$ at the same energy. Note that, although we are considering $\Kdf$ only in the isotropic approximation, the three-to-three scattering amplitude still depends on the incoming and outgoing three-particle phase space, indicated here with the shorthand $\Omega_3 \equiv (\vec k, \hat a^*)$.
The primary motivation of this work is to demonstrate the practical utility of our result.
Thus, for the sake simplicity, we consider only the $(\vec P=0)$-frame.
This allow us to
use $E$ rather than $E^*$ to denote the simultaneous finite-volume and c.m.-frame energy. 
In the same spirit, and following Ref.~\cite{\Akakib},
we take $\cK_2^s$ to be given by the leading-order term
in the threshold expansion, i.e. the term involving the scattering length $a$. 

The expression for $F_3^\iso$ with $\vec P=0$ is
\begin{equation}
F_3^\iso(E, L) = \langle \textbf{1}|F_3^s|\textbf{1}\rangle=\sum_{k,p} \left[ F_3^s \right]_{kp}
\,,
\label{eq:F3iso}
\end{equation}
where $|\textbf{1}\rangle$ has been defined above,
and the sum over the momenta ${k,p}$ is of finite range because $F_3^s$ is 
truncated by the cutoff function, $H(\vec k)$. 
Here and below we keep dependence on $E$ and $L$ implicit.
The matrix $F_3^s$ is given by
\begin{equation}
\left[F_3^s\right]_{kp}=  \frac{1}{L^3}\left[ \frac{\wt F^s}3 - \wt F^s
\frac1{1/(2\omega \cK_2^s) + \wt F^s + \wt G^s} \ \wt F^s\right]_{kp} \,,
\label{eq:F3s}
\end{equation}
where 
\begin{align}
\left[\frac1{2 \omega \cK_2^s}\right]_{k p}  &
=  \delta_{kp}
\left\{ - \frac1a   + |q_{2,k}^*|[1-H(\vec k)] \right\}
\frac1{32 \pi \omega_k E_{2,k}^*}
\,,
\label{eq:K2s}
\\
\widetilde F^s_{kp} &= \delta_{k p} \widetilde F^s(\vec k)\,,
\label{eq:Fs}
\\
\widetilde F^s(\vec k) &= \frac{H(\vec k)}{4\omega_k}
 \left[\frac1{L^3}\sum_{\vec a}^{\text{UV}} - \PV\int_{\vec a}^{\text{UV}} \right]
\frac{1}
{4\omega_a \omega_{ka} (E-\omega_k-\omega_a-\omega_{ka})}\,,
\label{eq:Fsa}
\\
\widetilde G^s_{kp} &= \frac{H(\vec k) H(\vec p)}{8 L^3 \omega_k \omega_p \omega_{kp}
(E-\omega_k-\omega_p-\omega_{kp})}
\,.
\label{eq:Gs}
\end{align} 
Here $\omega_k$ and $\omega_{ka}$ are the on-shell energies
for particles with momenta $\vec k$ and $\vec k+\vec a$, respectively, i.e.
\begin{equation}
\omega_k=\sqrt{\vec k^2+m^2}\,,\ \
\omega_{ka}=\sqrt{(\vec k+\vec a)^2+m^2}\,.
\end{equation}
Other $\omega$s are defined analogously.
The two-particle c.m.~energy and relative momentum are given by
\begin{align}
E_{2,k}^{*\, 2} &= (E-\omega_k)^2- \vec k^2 = E^2+m^2 - 2 E \omega_k\,,
\\
q_{2,k}^{*\, 2} & = {E_{2,k}^{*\,2}}/{4} - m^2\,.
\end{align}
The sum over $\vec a$ in Eq.~(\ref{eq:Fsa}) runs over all finite-volume momenta, while 
the integral is defined as $\int_{\vec a} \equiv \int d^3a/(2\pi)^3$.
The principal value (PV) prescription is defined such that the integral is an analytic
function of $\vec k^{\, 2}$ (and is referred to in Ref.~\cite{\HSQCa} as the
$\wt{\PV}$ prescription). Finally, the cutoff function is\footnote{%
Note that, for $\vec P=0$, $H(\vec k)=H(k)$. Nevertheless we keep the more general
notation for consistency with Refs.~\cite{\HSQCa,\HSQCb,\BHSQC} and because
$H$ does depend on $\vec k$ when $\vec P\ne 0$.}
\begin{align}
\label{eq:Hdef}
H(\vec k) & = J(z)\,, 
\\
z&= \frac{E_{2,k}^{*2} - (1+\alpha) m^2}{(3-\alpha)m^2} \,,
\\
J(z) &= 
\begin{cases}
0 \,, & z \le 0 \,; \\ \exp \left( - \frac{1}{z} \exp \left
[-\frac{1}{1-z} \right] \right ) \,, & 0<z < 1 \,; 
\\ 1 \,, & 1\le z
\,.
\end{cases}
\end{align}
This is the form introduced in Ref.~\cite{\BHSQC}, chosen to smoothly
interpolate between $0$ and $1$ as $E_{2,k}^*/m$ ranges from
$\sqrt{1+\alpha} $ to the threshold value of $2$.
In the following we consider $\alpha=-1$, which gives the maximum allowed range.\footnote{%
The relationship between $\alpha$ and the parameter $\delta$ used earlier in this section
is $\sqrt{1+\alpha}=2-\delta$. Thus $\alpha=-1$ corresponds to $\delta=2$.
}

In Eq.~(\ref{eq:Fsa}) we have labeled both the sum and the integral with a superscript ``UV'' indicating that an ultraviolet cutoff is required to separately evaluate the sum and integral. In Refs.~\cite{\HSQCa,\HSQCb} a specific choice of cutoff is used, namely the product of two of the smooth cutoff functions, $H(\vec a) H(-\vec a-\vec k)$. We primarily use this definition in this work as well, but we also make use of
the definition given in Ref.~\cite{Kim:2005gf} for some quantities.
These two definitions are described in more detail in Appendix~\ref{app:Fs}, 
where we also explain our method of numerical evaluation.
In places where we use both definitions, we refer to that using $H$-functions for the UV cutoff
as $\widetilde F^s_{\rm HS}$, and that using the approach of Ref.~\cite{Kim:2005gf} as
$\widetilde F^s_{\rm KSS}$. 
The subscripts abbreviate the authors of the article where each cutoff was first introduced.

It is important to note that the freedom to adjust the ultraviolet cutoff here is logically separate from 
the freedom in the choice of $H(\vec k)$ in Eqs.~(\ref{eq:K2s}) and (\ref{eq:Gs}). 
Varying the UV regulator in Eq.~(\ref{eq:Fsa}) changes the value of $\widetilde F^s$
only by the exponentially suppressed corrections 
that we are ignoring throughout.\footnote{%
Strictly speaking, this holds only if the regulator  only modifies the terms satisfying 
$ \vert E - \omega_k - \omega_a - \omega_{ka}\vert \gg m$, 
and equals unity when $  E - \omega_k - \omega_a - \omega_{ka}=0$.}
Thus we can choose the regulator that is most convenient for numerical evaluation.
By contrast, varying factors of $H(\vec k)$ outside a sum-integral difference,
such as in Eqs.~(\ref{eq:K2s}) and (\ref{eq:Gs}), leads to changes $F_3^s$ that are,
in general, not exponentially suppressed. 
These are such, however,  that $\Kdf$ can in principal be adjusted to keep the low-energy
physics unchanged. In other words, an adjustment in the external $H$ functions corresponds to a 
change in the renormalization scheme.\footnote{%
Despite this expectation, 
we discuss below examples where exponentially suppressed
finite-volume artifacts can lead to significant effects,
e.g.~the unphysical solutions discussed in Sec.~\ref{sec:unphys}.
 }

The form of the result for $1/\cK_2^s$ in Eq.~(\ref{eq:K2s}) deserves further explication.
Above threshold, where $H(\vec k)=1$, this form arises from the standard s-wave K matrix, 
$16 \pi E_{2,k}^* \tan \delta_0(q^*_{2,k})/q^*_{2,k}$,
keeping only the leading order in the threshold expansion. 
Below threshold, the result interpolates smoothly to 
the subthreshold s-wave scattering amplitude, $\cM_2^s$,
reaching this amplitude when $H(\vec k)\to 0$.
As explained in Ref.~\cite{\HSQCa}, this behavior follows from the choice of
pole prescription in $\wt F^s$.
\begin{comment}
Here we recall that $\cK_2^{s,\,\rm std}$ and $\cM_2^s$ are related by
\begin{align}
\frac1{K_2^{s,\,\rm std}(\vec k)} &= \frac1{\cM_2^s(\vec k)} + \rho(\vec k) \,,
\\
\rho(\vec k) &=  \frac{1}{16 \pi E_{2,k}^*}
\begin{cases}
-i q_{2,k}^* & E_{2,k}^{* 2} \ge 4 m^2 \,;
\\
|q_{2,k}^*| & E_{2,k}^{* 2} < 4 m^2 \,.
\end{cases}
\end{align}
\end{comment}

\bigskip

From these definitions we see that $F_3^\iso$ depends on $E$, $L$ and $a$.
For fixed $L$ and $a$, the spectrum is determined by those
values of $E$ for which $F_3^\iso(E) \Kiso(E) = -1$. In Appendix~\ref{app:numerics}, we 
describe how we implement this numerically.
Here we note two caveats. First, the formalism breaks down as $E$ approaches $5m$, where 
the five-particle channel becomes important. 
Second, the formalism does not hold if $\cK_2^s$ has a pole in the region
of $E_{2,k}^*$ that enters into the calculation, namely
$\sqrt{1+\alpha} < E_{2,k}^*/m < (E^* - m)/m  $. Note that this restriction includes poles
below as well as above threshold.

With the form of $\cK_2^s$ that we use,
Eq.~(\ref{eq:K2s}), we see that there are no poles above threshold, but there is
a pole below threshold if 
\begin{equation}
1/a = {|q_{2,k}^*| [1 - H(\vec k)]}
\,.
\end{equation}
One can show that the right-hand side lies between $0$ and $m$ for all
allowed values of $E$, $\vec k$ and $\alpha$.
Thus to avoid the poles in general the scattering length must satisfy\footnote{%
In fact, for $\alpha > -1$, a somewhat higher, $\alpha$-dependent upper limit applies.}
\begin{equation} 
a < 1/m \,.
\end{equation}
We stress that negative values of $a$ having arbitrarily large magnitude are allowed,
so we can investigate the unitary limit. 
Indeed, as can be seen from Eqs.~(\ref{eq:K2s}) and (\ref{eq:F3sa}), we can
work directly at $1/a=0$, although we do not make use of this possibility in our
numerical studies.

\subsection{Relation between $\Kiso$ and $\mathcal M_3$\label{sec:KtoM}}

We close this section by recalling from Ref.~\cite{\HSQCb} the relation between 
the infinite-volume quantities $\Kdf$ and $\cM_3$. 
In the isotropic approximation, this requires solving only one integral equation.
This is for the quantity $\cD^{(u,u)}(\vec k, \vec p)$
that sums up repeated two-particle scattering in which the two particles involved
can switch any number of times.
It satisfies
\begin{equation}
\cD^{(u,u)}(\vec k, \vec p) = - \cM_2^s(\vec k) G^\infty(\vec k, \vec p) \cM_2^s(\vec p)
- \int_{\vec s} \frac1{2\omega_s} \cM_2^s(\vec k) G^\infty(\vec k, \vec s) \cD^{(u,u)}(\vec s,\vec p)
\,,
\label{eq:Duu}
\end{equation}
where, as usual, $\vec k$ and $\vec p$ are spectator momenta, which are
now continuous variables.
$\cM_2^s(\vec k)$ is the physical s-wave two-particle scattering amplitude
with two-particle c.m.~energy $E_{2,k}^*$, 
which in the low-energy approximation is given by
\begin{align}
\frac1{\cM_2^s(\vec k)} &=  - \frac1a \frac1{16 \pi E_{2,k}^*} + \rho(\vec k)
\,,
\label{eq:M2s}
\\
\rho(\vec k) &=  \frac{1}{16 \pi E_{2,k}^*}
\begin{cases}
-i q_{2,k}^* & E_{2,k}^{* 2} \ge 4 m^2 \,;
\\
|q_{2,k}^*| & E_{2,k}^{* 2} < 4 m^2 \,,
\end{cases}
\label{eq:rho}
\end{align}
and $G^\infty$ is an infinite-volume quantity related to $\wt G^s$,
\begin{equation}
G^\infty(\vec k, \vec p) = \frac{H(\vec k) H(\vec p)}{2\omega_{kp} 
(E-\omega_k-\omega_p-\omega_{kp} + i \epsilon)}
\,.
\label{eq:Ginf}
\end{equation}
The cutoff functions imply that the integral is of finite range. 
Note that we are using the $i\epsilon$ pole prescription here. This is correlated
with the appearance of the scattering amplitude $\cM_2^s$, rather than
$\cK_2^s$, in the integral equation.

Above threshold, $\cM_3$ has singularities at particular, physical kinematic points,
and so in Ref.~\cite{\HSQCa} we introduced a divergence-free version of the amplitude
\begin{equation}
\Mdf(\vec k, \hat a^*;\vec p, \hat a'^*) = 
\cM_3(\vec k, \hat a^*;\vec p, \hat a'^*) - \cS\left\{\mathcal D^{(u,u)}(\vec k, \vec p)\right\}
\,.
\label{eq:Mdf}
\end{equation}
The notation here is that $\cS$ is a symmetrization operator that sums over the
three choices of spectator momentum for both initial and final states.
The need for such symmetrization
implies that $\Mdf$ and $\cM_3$ depend not only on the spectator momenta,
but also on the directions of the other two particles in their relative c.m.~frame,
which are given by $\hat a^*$ and $\hat a'^*$ respectively for the initial and final states.
$\Mdf$ has the advantage compared to $\cM_3$
of being a smooth function of momenta and $E$, 
so that, in particular, it is well defined at threshold.
It has the disadvantage of depending on the cutoff function $H$.

In the isotropic approximation, $\Mdf$ is related to $\Kiso$ by
\begin{align}
\Mdf (\vec k, \hat a^*;\vec p, \hat a'^*) &= 
\cS\left\{\cL(\vec k) \frac1{1/\Kiso + F_3^\infty} \cR(\vec p) \right\}\,,
\label{eq:M3dKiso}
\end{align}
where
\begin{align}
\cR(\vec k) & = \cL(\vec k) = \tfrac13 - 2 \omega_k \cM_2^s(\vec k) \wt \rho(\vec k) 
- \int_{\vec s} \cD^{(u,u)}(\vec k,\vec s)
\wt \rho(\vec s)\,,
\label{eq:cL}
\\
F_3^\infty &= \int_{\vec k} \wt \rho(\vec k) \cL(\vec k)\,,
\label{eq:F3inf}
\\
\wt \rho(\vec k) &= \frac{H(\vec k) \rho(\vec k)}{2\omega_k} \,.
\label{eq:wtrho}
\end{align}
These relations involve only integrals over a finite range of momenta.
In Appendix~\ref{app:numerics}, we discuss how these quantities can be readily determined at or below threshold.

\section{Numerical Results \label{sec:results}}

In this section we present a sampling of numerical results, 
aiming both to provide checks by comparing with several known analytic results,
and to give examples of the finite-volume spectrum that emerges for various
choices of the scattering parameters.
Throughout this section we use units in which $m=1$, with the exception of
the figures, where for clarity we add back in appropriate factors of $m$. 
Most of the details regarding the numerical evaluation of the 
finite-volume functions are described in the appendices.

\subsection{Energy spectrum with $\Kiso=\Mdf=0$}
\label{sec:Kisozero}

We begin by studying the finite-volume spectrum for the special case
of $\Kiso=0$. This in turn implies that $\Mdf=0$ 
and thus that  $\cM_3 = \cS \, \big \{\mathcal D^{(u,u)}(\vec k, \vec p\,) \big \}$
[see Eqs.~(\ref{eq:Mdf}) and (\ref{eq:M3dKiso})].
In words this says that
the three-to-three scattering amplitude is given by the sum
over all pair-wise scattering diagrams in which the two-particle
subprocesses are mediated by on-shell two-to-two scattering
amplitudes.\footnote{%
We stress that both $\Kdf$ and $\Mdf$ are scheme
dependent in the sense that physical predictions for a given $\Kdf$
can only be made once a particular form of $H$ has been specified. 
Thus, when we say that $\Kiso=\Mdf=0$, this is for $H$ defined in
Eq.~(\ref{eq:Hdef}) with $\alpha=-1$.}
In this case the quantization condition simply becomes
\begin{equation}
1/F_3^\iso(E, L, a) = 0 \,.
\label{eq:QCisoK0}
\end{equation}
This is a useful starting point because it provides a benchmark for
three-particle lattice calculations. 
If three-particle energies were
found to be consistent with the $\Kiso=0$ predictions, then it would only be
possible to place upper limits on $\Kiso$.
By contrast, resolving a shift from these values would gives a
direct indication of the strength of this local three-body interaction.
The solutions to Eq.~(\ref{eq:QCisoK0}) occur at the poles in $F_3^\iso$.
The numerical determination of the positions of these poles is straightforward, 
as described in Appendix~\ref{app:numQC}.
Examples of the form of $F_3^\iso$ are shown in Figs.~\ref{fig:QCexample} and \ref{fig:F3isoE4L10}.

\begin{figure}
\begin{center}
\includegraphics[width=0.75\textwidth]{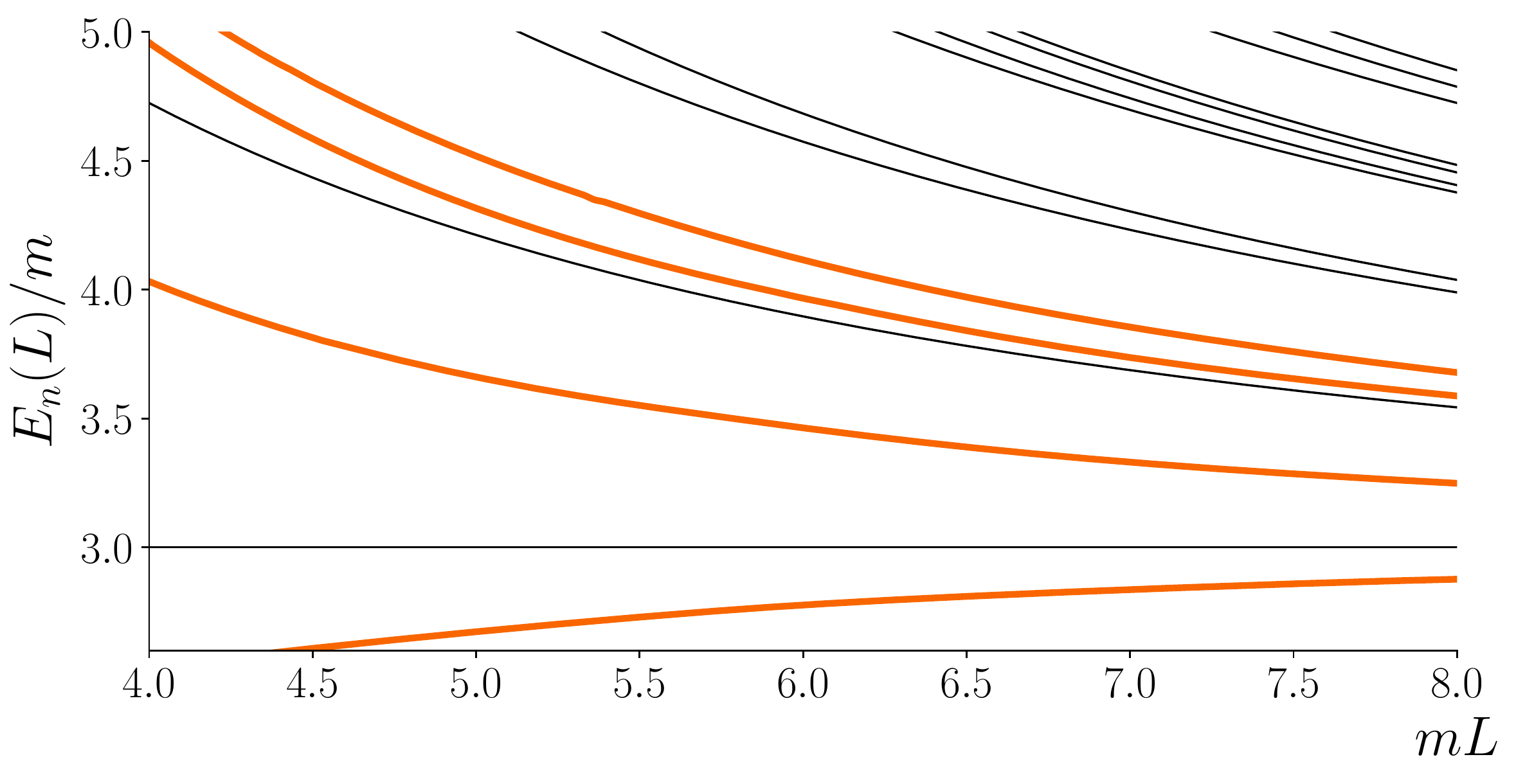}
\caption{The four lowest-energy solutions to the quantization condition for $\Kiso=0$ and $ma=-10$ 
(thick, orange) together with the lowest eleven noninteracting levels, 
i.e.~solutions for $\Kiso=a=0$, (thin, black). 
We only show results in the energy range for which our formalism is valid, namely
$E_n < 5m$.
Noninteracting levels are clustered according to momenta that are degenerate in the
 nonrelativistic theory, as discussed in the text. 
}
\label{fig:excitedKdf0}
\end{center}
\end{figure}

In Fig.~\ref{fig:excitedKdf0} we plot the low-lying finite-volume spectrum
for $\Kiso=0$ and $a=-10$, together with the noninteracting three-particle levels.
The latter are given by
\begin{equation}
E_n(L) = \sqrt{1 + (2 \pi/L)^2 \vec m_1^2} + 
\sqrt{1 + (2 \pi/L)^2 \vec m_2^2} + 
\sqrt{1 + (2 \pi/L)^2 \vec m_{12}^2} \,,
\end{equation}
where $\vec m_1$ and $\vec m_2$ are integer vectors 
determining the momentum of two of the particles, 
while $\vec m_{12}=-\vec m_1-\vec m_2$ determines the momentum of the third.
In Table~\ref{tab} we collect some information about
the low lying noninteracting levels.
The values of $L$ that are shown correspond to those
used in present lattice QCD simulations ($ 4 \lesssim m_\pi L \lesssim 6$)
as well as somewhat larger values that may be accessible in the future.

\begin{table}
\caption{
Summary of noninteracting three-particle energies, in units where the particle mass is $m=1$. 
The first column gives the index of the level, ordered by increasing energy for large $L$ 
(which means $L \gtrsim 9.5$ for these levels). 
The second column gives the three squared integers describing the individual momenta.
The third column gives the degeneracy for identical particles.
The final two columns give the energies for  $L=4$, $6$, and $10$.
Horizontal lines group levels having the same value of  the sum
$\vec m_1^2 + \vec m_2^2 + \vec m_{12}^2$, which are thus degenerate in the 
nonrelativistic limit.
We show all the levels having  values of this sum up to 12.
\label{tab}}
\vspace{10pt}
\begin{tabular}{c || c  || c || c || c || c}
\ \ \ $n$ \ \ \  & $\left(\vec m_1^2,\vec m_2^2, \vec m_{12}^2\right)$ & \ \ \ degeneracy\ \ \  & \ \ \  $E_n(L=4)$ \ \ \  & \ \ \ $E_n(L=6)$ \ \ \ & \ \ \  $E_n(L=10)$ \\ \hline\hline
1 & (0 , 0 , 0) & 1 & 3.0 & 3.0 & 3.0 \\ \hline
2 & (1 , 1 , 0) & 3 & 4.72 & 3.90 & 3.36\\ \hline
3 & (2 , 2 , 0) & 6 & 5.87 & 4.57 & 3.68\\
4 & (2 , 1 , 1) & 12 & 6.16 & 4.68 & 3.70\\ \hline
5 & (3 , 3 , 0) & 4 & 6.80 & 5.14 & 3.96\\
6 & (4 , 1 , 1) & 3 & 7.02 & 5.22 & 3.97\\
7 & (3 , 2 , 1) & 24 & 7.20 & 5.31 & 4.00\\
8 & (2 , 2 , 2) & 8 & 7.31 & 5.36 & 4.01\\ \hline
9 & (4 , 4 , 0) & 3 & 7.59 &  5.64 & 4.21\\
10 & (5 , 2 , 1) & 24 & 7.95 & 5.78 & 4.24\\
11 & (4 , 2 , 2) & 12 & 8.17 & 5.89 & 4.28\\ \hline
12 & (5 , 5 , 0) & 12 & 8.30 & 6.09 & 4.45\\
13 & (6 , 3 , 1) & 24 & 8.74 & 6.27 & 4.49\\
14 & (6 , 2 , 2) & 24 & 8.85 & 6.33 & 4.51\\
15 & (5 , 4 , 1) & 24 & 8.81 & 6.32 & 4.51\\
16 & (5 , 3 , 2) & 48 & 8.99 & 6.40 & 4.54\\
17 & (4 , 3 , 3) & 12 & 9.09 & 6.46 & 4.56\\ \hline
18 & (6 , 6 , 0) & 12 & 8.95 & 6.51 & 4.67\\
19 & (8 , 2 , 2) & 6 & 9.43 & 6.70 & 4.71\\
20 & (6 , 5 , 1) & 48 & 9.49 & 6.75 & 4.74\\
21 & (6 , 4 , 2) & 24 & 9.71 & 6.86 & 4.78\\
22 & (5 , 5 , 2) & 36 & 9.74 & 6.88 & 4.79\\ \hline
\end{tabular}

\end{table}

\begin{figure}
\begin{center}
\includegraphics[width=\textwidth]{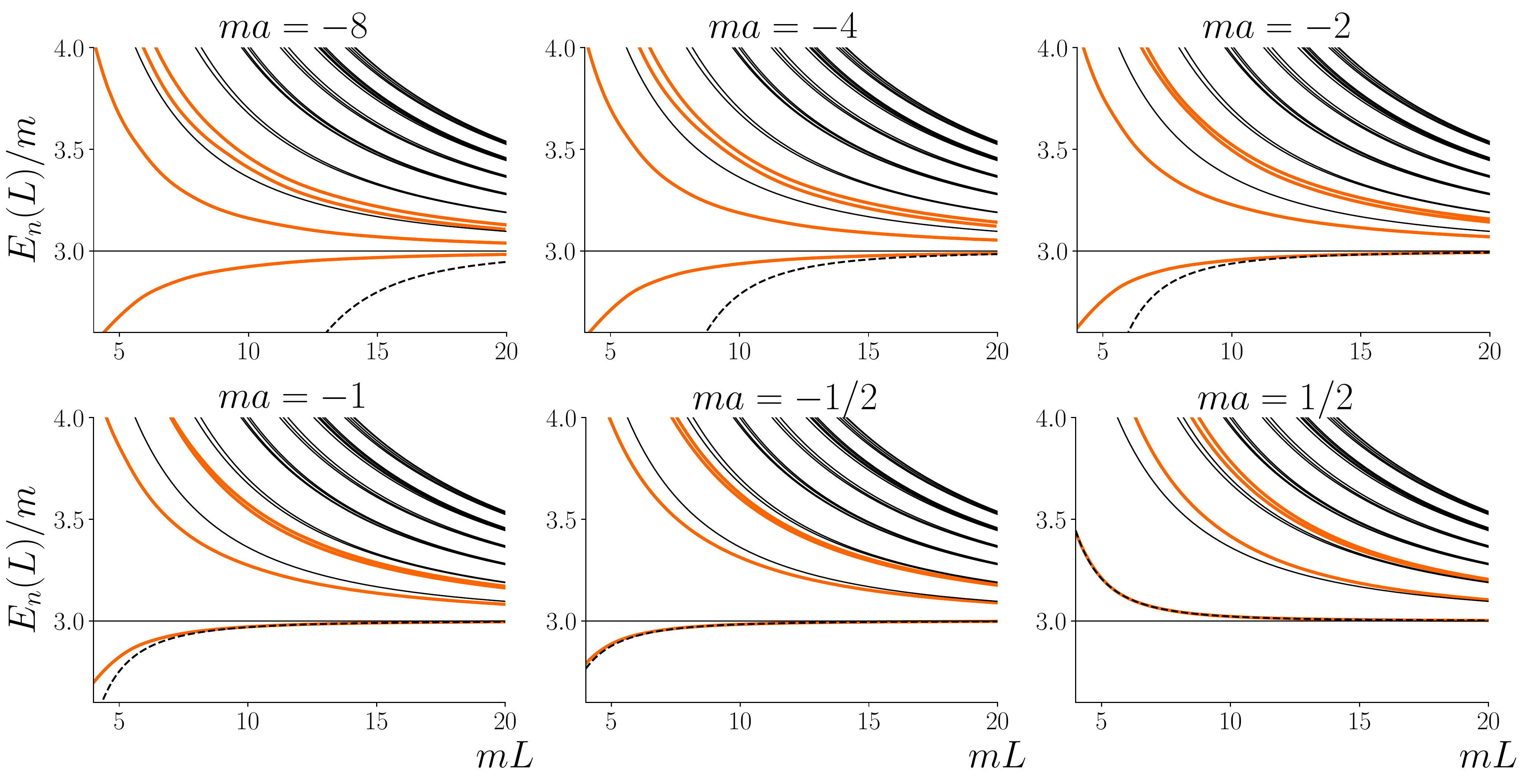}
\caption{$E_n(L)/m$ vs $mL$ for $\Kiso=0$ and various values of the scattering length, $a$. 
Notation as in Fig.~\ref{fig:excitedKdf0}, although a larger range of $mL$ is displayed here,
as well as additional noninteracting levels.
The dashed black curve shows the threshold expansion, Eq.~(\ref{eq:threxp}) 
through $\mathcal O(1/L^5)$.
}
\label{fig:EvsLVaryA}
\end{center}
\end{figure}

{
Our interpretation of the $a=-10$ levels is that they 
correspond to the first four noninteracting levels, but pushed to significantly lower
energies by the  strongly attractive two-particle interaction.
In particular, the lowest state is not a bound state.
We can see this by extending the calculation to larger values of $L$ and observing
that it approaches the threshold energy $E=3$.
We refer to this state below as the threshold state.

We have shown several additional noninteracting levels in Fig.~\ref{fig:excitedKdf0}
in order to illustrate the clustering of excited states.
This clustering can be understood by doing a nonrelativistic expansion of the energies.
In particular, keeping only the leading term---that present in nonrelativistic quantum
mechanics (NRQM)---the
noninteracting energies are 
\begin{equation}
E^{\text{n.r.}}_n(L) = 3 + \frac{2 \pi^2}{L^2} \big (\vec m_1^2 +  \vec m_2^2 + \vec m_{12}^2 \big ) \,.
\end{equation}
As a result, all states for which the sum of squared momenta are equal become degenerate.
This increased degeneracy is indicated by the groupings in Table~\ref{tab}. 
The gaps within the clusters scale as
\begin{equation}
E_n(L) - E^{\text{n.r.}}_n(L) = -  \frac{2 \pi^4}{L^4}  \big (\vec m_1^4 +  \vec m_2^4 + \vec m_{12}^4 \big ) + \mathcal O(1/L^6) \,,
\end{equation} 
whereas the gaps between different clusters scale as $1/L^2$.\footnote{%
Interestingly, for the threshold state, the effect of interactions scales with
 the power between these two, i.e.~as $1/L^3$ (as discussed in detail below).}
We note that the splittings within clusters become significant 
for the values of $m L$ used in present simulations,
i.e.~those at the lower end of the range displayed.
This indicates the importance of including relativistic kinematics in order to
gain sufficient precision in the spectrum.
  
One issue that is potentially confusing concerns the degeneracies of the levels shown in 
Fig.~\ref{fig:excitedKdf0}. Solutions to the quantization condition in the isotropic approximation
are nondegenerate, whereas the noninteracting levels are highly degenerate, as can be seen
from Table~\ref{tab}. As explained in Ref.~\cite{\HSQCa}, the resolution is that, even in the
presence of interactions, all but one of the degenerate levels remain at the noninteracting
energy {\em when working in the isotropic approximation}. These remaining levels will be shifted and split upon inclusion of nonisotropic interactions.
We also note that one can project onto the states shown in the figure in practice
by using three-particle operators living in $A_1^+$ irreducible representation (irrep)
of the cubic group. This irrep has overlap with the state $|\mathbf{1}\rangle$,
and also picks out a single state from each of the noninteracting levels.

In Fig.~\ref{fig:EvsLVaryA} we show the
result of varying the scattering length. The upper left panel shows
$a=-8$, which is very similar to $a=-10$, while subsequent panels halve the value of $a$,
with the exception of the final panel, which shows the result for a small, positive $a$.
(We recall that the maximum value for which our formalism holds is $a=1$.)
In these figures we extend the range of $mL$ up to $20$, which allows one to see clearly
the approach of the levels to the noninteracting curves as $a\to 0$.
The larger range also allows us to show additional noninteracting levels, and thus further
emphasize the clustering discussed above.
Finally, we add to the plot the prediction for the energy of the threshold state
in an expansion in powers of $a/L$, Eq.~(\ref{eq:threxp}).
We observe that this expansion works well for the smallest values of $|a|$.
We investigate this expansion in more detail in Sec.~\ref{sec:threxp} below.

}

\subsection{Energy spectrum for nonzero $\Kiso$}
\label{sec:Kisononzero} 

\begin{figure}
\begin{center}
\includegraphics[width=\textwidth]{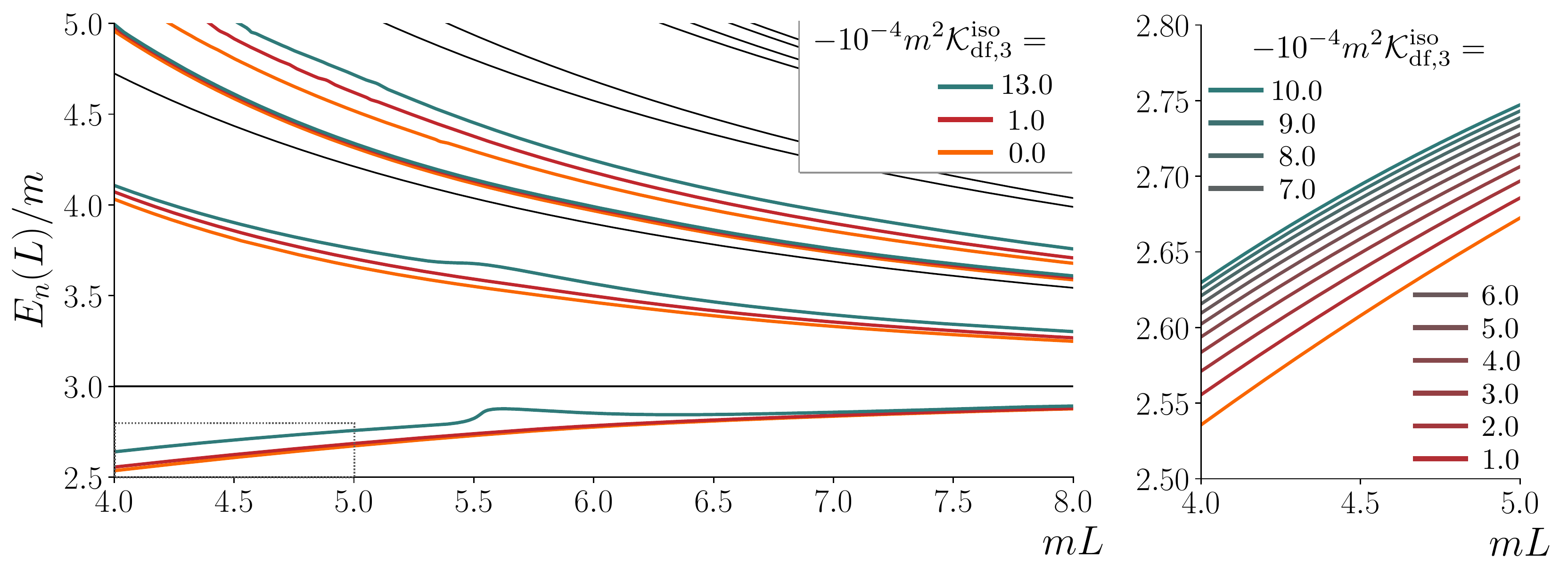}
\caption{
Finite-volume energy levels for $ma=-10$ and various negative values of $m^2 \Kiso$.
The left plot shows results from two nonzero values of $\Kiso$,
as well reproducing the $\Kiso=0$ results, and the noninteracting levels, 
from Fig.~\ref{fig:excitedKdf0}.
Note that the extent to which $\Kiso$ shifts the energy depends significantly on the level being considered. The right panel magnifies the region shown by the dashed rectangle in
the left panel, displaying results for the lowest energy state from a larger number of nonzero values of $\Kiso$.
\label{fig:negKdf}
}
\end{center}
\end{figure}

We now consider solutions to the quantization condition with nonzero  $\Kiso$. 
We first take energy-independent, negative values of $\Kiso$. 
As with the two-particle K matrix, small negative values of $\Kiso$ correspond to repulsive interactions, and thus push the levels up.
We illustrate this in Fig.~\ref{fig:negKdf} for the case of $a=-10$ shown previously for
$\Kiso=0$ in Fig.~\ref{fig:excitedKdf0}.
The levels increase monotonically as $\Kiso$ becomes more negative.
Large magnitudes of $\Kiso$ are required to see a noticeable shift because, as we discuss in more detail below, for small values of $\Kiso$ and $a$, the effect of the three-body contact interaction on the energy is suppressed by $1/L^6$. 
In this regard, we stress that such large values of $|\Kiso|$ are not unphysical.
Indeed, as can be seen from Eq.~(\ref{eq:M3dKiso}), the three-particle scattering amplitude
is finite in the $|\Kiso|\to \infty$ limit. 
This is analogous to the two particle sector where $\cK_2 \to \infty$ corresponds to 
the unitary limit, $\cM_2 = i 16 \pi E_{2}^*/q^*_{2}$.

One noticeable feature of Fig.~\ref{fig:negKdf} is the appearance of a ``bump" in the curves
around $L=5.5$. If $\Kiso$ is made even more negative the spectral lines double back,
which is an unphysical result. We discuss this issue further in Sec.~\ref{sec:unphys}.
What we want to stress here is that, for most values of $\Kiso$, $a$ and $L$, the 
quantization condition in the isotropic approximation gives reasonable results,
with energy levels that are sensitive to the three-particle interaction.

\begin{figure}
\begin{center}
\includegraphics[width=\textwidth]{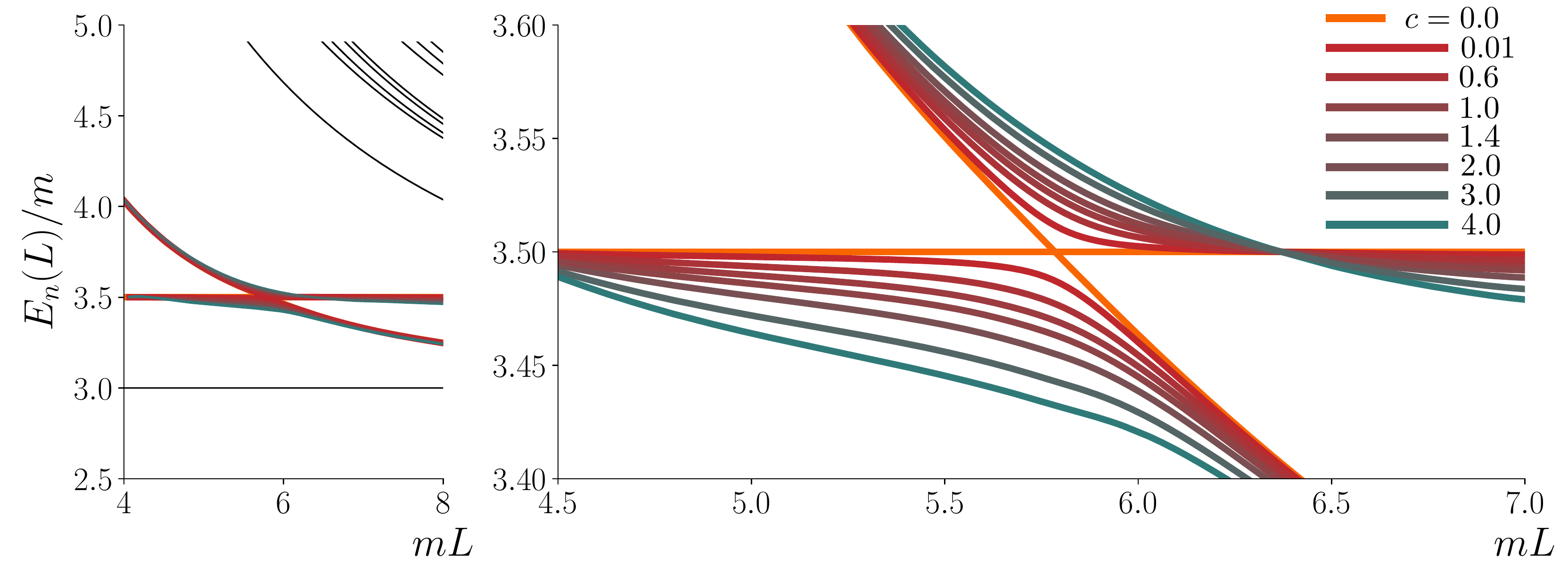}
\caption{Finite-volume energies for $ma=-10$ and a ``Breit-Wigner" ansatz for $\Kiso$, Eq.~(\ref{eq:BWansatz}). The constant $c$ characterizes the strength of coupling
between the resonance of mass $M_R = 3.5 m$ and the scattering states. For nonzero $c$
the crossing of $E(L) = M_R$ with the scattering states is replaced with an avoided level crossing. 
{The left panel gives an overview of the position of the crossing in the overall spectrum, 
while the right panel zooms in on the crossing itself.}
\label{fig:res}
}
\end{center}
\end{figure}

A more striking example of this sensitivity is shown in Fig.~\ref{fig:res},
where we use the freedom to allow $\Kiso$ to depend on energy to model a
three-particle resonance. 
The ansatz we use is
\begin{equation}
\label{eq:BWansatz}
\Kiso(E) = - \frac{c \times  10^3}{E^2 - M_R^2} \,,
\end{equation}
with a ``resonance mass" of $M_R = 3.5$. 
This form is inspired by the standard Breit-Wigner parametrization of the two-particle K matrix, although further investigation is needed to understand if this gives a physical 
description of three-particle resonances. 
At the very least, however, it gives a unitary description of three-to-three scattering that, 
as $c \to 0$, smoothly deforms to a decoupled system of a stable state with mass $M_R$ 
together with three-particle scattering states. For nonzero values of $c$ the two sectors couple and the avoided-level crossings characteristic of a resonance are observed, with
the gap increasing with $c$.

For a physical system described by this ansatz, fitting lattice-determined finite-volume levels would give constraints on $c$, $M_R$ and the scattering length $a$. Consideration of how this ansatz for $\Kiso$ converts to $\mathcal M_3$, and whether this gives a useful three-particle resonance description, is a topic for future study.

\subsection{Volume-dependence of the energy of a bound state}
 \label{sec:bound}

\begin{figure}
\begin{center}
\includegraphics[width=\textwidth]{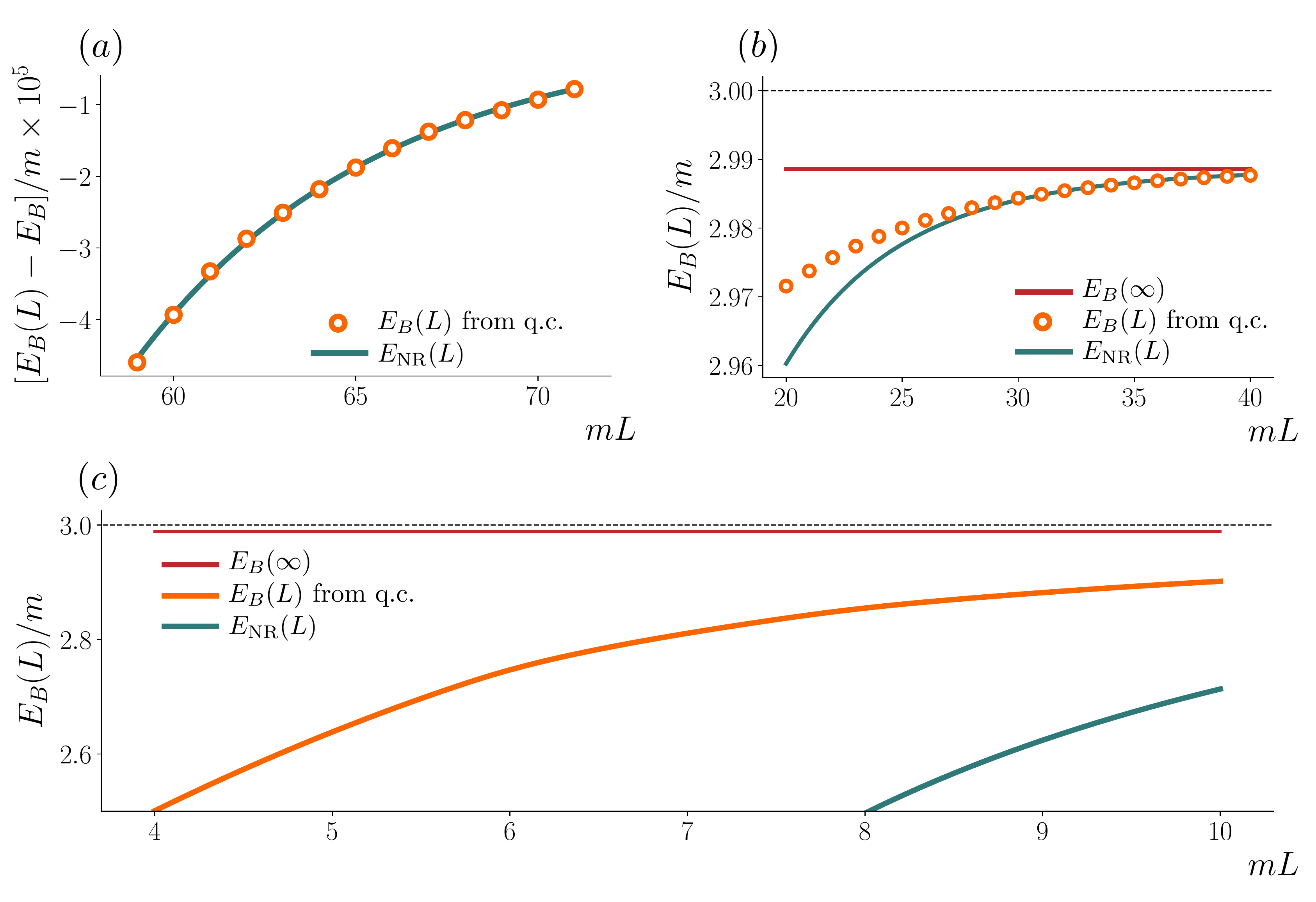}
\caption{Finite-volume energy dependence for the bound state that arises for $m^2 \Kiso=2500$ and $ma=-10^4$. 
In all three figures the solutions to the quantization condition are marked in orange, as points in (a) and (b) and as the curved solid line in (c). The curving (turquoise) line in panel (a) is a fit of 
Eq.~(\ref{eq:FVBS}) (neglecting the higher-order corrections) 
to the data in this panel. The same fit line is shown in panel (b)
for lower values of $mL$, along with
a horizontal, solid (red) line showing the infinite-volume energy of the bound state $E_B(\infty)$.
The horizontal dashed (black) line shows the threshold energy $E=3m$.
Panel (c) displays $E_B(L)$ for smaller m$L$, along with the same two horizontal lines as in (b) and the asymptotic prediction.
\label{fig:BS1}
}
\end{center}
\end{figure}

In this section we provide a quantitative test of our numerical results by studying
the volume dependence of the energy of a bound state $E_B(L)$ in the unitary regime, 
$|a| \gg 1$. This can be compared with the analytic result of Ref.~\cite{\AkakiBS},
\begin{equation}
E_{B}(L) = 3 - \kappa^2 - 96.35 |A|^2 \kappa^2 \frac{e^{- 2\kappa L/\sqrt{3}}}{(\kappa L)^{3/2}} 
\left[1 + \mathcal O \left (\frac{1}{\kappa L} , e^{- \alpha \kappa L}, \kappa^2\right ) 
\right ]
\,,
\label{eq:FVBS}
\end{equation}
where $\kappa$ is defined in terms of the
infinite-volume value of the bound state energy, $E_B(\infty)=3-\kappa^2$,
 $|A|^2$ is a normalization factor that is expected to be close to unity,\footnote{%
For a detailed discussion of the significance of $A$ see Ref.~\cite{\Akakia}.}
while $\alpha$ is of $\mathcal O(1)$.
This result is valid if $\kappa \ll 1$ (nonrelativistic regime), $|a| \gg 1$ (two-particle
unitary regime), and $ \kappa L\gg1$.
In addition, two-particle interactions are assumed to be s-wave dominated, and the
bound state is assumed to have $J=0$.

We note that this result can also be derived analytically from our quantization condition,
under the same assumptions~\cite{\HSBS}.
This derivation applies also within the low-energy isotropic approximation.
However, this derivation requires crucial external input beyond the quantization
condition itself, namely 
the long-distance part of the Schr\"odinger wavefunction in the three-body system.
Thus agreement with Eq.~(\ref{eq:FVBS}) tests not only our numerical
methods, but also that the quantization condition itself correctly reproduces
the physics of the bound state.
We can also learn where the formula breaks down,
i.e.~where subleading volume-dependence enters.

With this in mind, we have numerically determined the bound-state energy for the parameters $a=-10^4$ (assuring that we are in the unitary regime) and $\Kiso = 2500$. 
Note that, in contrast to the previous section, here we choose $\Kiso$ positive,
as we find that this generically produces a bound state.\footnote{%
{Why this is the case will become clear in the following section. 
This result is analogous to the
fact that, in the two-particle case, a bound state occurs when $a$ is large and positive.}}
The results are shown in Fig.~\ref{fig:BS1}. We find that for $\kappa L>4$ ($L >  37$) 
$E_B(L)$ is well described by the asymptotic form given in Eq.~(\ref{eq:FVBS}). To be
conservative we do our final fit only to data for $L>59$ (corresponding to $\kappa L > 6.3$),
as shown in Fig.~\ref{fig:BS1}(a).
The fit gives $\kappa = 0.106844$, corresponding to a binding energy of $E_B = 2.98858$.
In addition we find $|A|^2=0.948$, and the fact that this result lies close to unity is a strong check on the applicability of the asymptotic form.

Figure.~\ref{fig:BS1}(b) compares the spectrum to the fit for smaller volumes, 
$20 < L < 40$. None of the data shown in this figure are used in the fit,
so the good qualitative agreement for $L>35$ provides a strong check that the result of
Eq.~(\ref{eq:FVBS}) is consistent with our quantization condition over a wide range of volumes. The deviation as one drops below $L \approx 30$ is also expected since $\kappa L$ then becomes too small and the asymptotic form no longer holds. 
We stress, however, that the solution to the quantization condition continues to be valid for all volumes shown, including the lowest range, $4<L<10$, shown in Fig.~\ref{fig:BS1}(c).
For smaller volumes the exponentially suppressed corrections that we are ignoring would
start to become sizable.

These results illustrate the potential utility of the quantization condition 
for analyzing three-particle bound-states. 
Given the value of $a$ from two-particle scattering, one can constrain $\Kiso$ near threshold using multiple three-particle scattering states. 
Extrapolating the results for $\Kiso$ to subthreshold energies, one can use
the quantization condition to predict the volume dependence of the bound state.
We see from Fig.~\ref{fig:BS1}(c) that, in the regime of $m L$ accessible to simulations,
the finite-volume energy shifts are large, and the asymptotic formula does not hold.
Thus the full quantization condition is needed to remove the finite-volume shift and
determine the infinite-volume binding energy. 
We also stress that, in this regime,
the bound-state energy is pushed so far below threshold that relativistic momenta are sampled. Thus a relativistic formalism is required to reliably describe even the near threshold state.

\subsection{Volume-dependence of the threshold-state energy}
\label{sec:threxp}

In this section we investigate in detail the energy of the threshold state.
We have already shown examples of this energy for various values of $a$
in Fig.~\ref{fig:EvsLVaryA}, and our aim here is to provide a detailed comparison with
the predicted large-volume behavior. The analytic prediction is
\begin{align}
E(L)- 3 &= \frac{c_3}{L^3} +\frac{c_4}{L^4}  +\frac{c_5}{L^5}
+ \frac{\tilde c_6}{L^6} -
\frac{\Mthr}{48 L^6} + \cO\left(\frac1{L^7}\right)
\,,
\label{eq:threxp}
\end{align}
where the coefficients are (using the fact that, in our approximation, the effective
range, $r$, vanishes)
\begin{align}
c_3 &={12\pi a }\,,
\\
\frac{c_4 }{c_3} &=  - \frac{a }{\pi} \cI\,,
\\
\frac{c_5}{c_3} &=  \left(\frac{a }{\pi}\right)^2 (\cI^2+ \cJ)\,,
\\
\frac{\tilde c_6}{c_3}&= \left(\frac{a }{\pi}\right)^3 \left[ -\cI^3+\cI \cJ +15 \cK + \cC_F+ \cC_4+\cC_5
+ \frac{16 \pi^3}3 (3 \sqrt3 -4\pi) \log\left(\frac{L}{2\pi}\right)\right]
+{64 \pi^2 a^2} \cC_3 + {3\pi a} \,.
\end{align}
The numerical values of the constants entering these expressions are\footnote{%
Note that we need $\cI$ to greatest accuracy, followed by $\cJ$, while $\cK$, $\cC_3$ etc.~are needed to lower accuracy.}
\begin{gather}
\cI= -8.913 632 917 59\,,\ \
\cJ= 16.532315960\,,\ \
\cK=8.401 923 974 828\,,\ \
\\
\cC_3 = -0.05806\,,\ \
\cC_F+\cC_4+\cC_5=2052\,.
%\cC_F = - 0.493036\,,\ \  \cC_4 = 105.892\,,\ \ \cC_5 = 1947\,.
\end{gather}
The terms through $\cO(1/L^5)$ were derived in NRQM in Refs.~\cite{Beane2007,Tan2007}.
Relativistic effects first enter at $\cO(1/L^6)$, and the relativistic form of $\tilde c_6$
was determined in Ref.~\cite{\HSTH} from our three-particle quantization condition.
The derivation was done including all partial waves in $\cK_2$ and $\Kdf$, 
but holds also in the isotropic limit.

Considering only terms through $\cO(1/L^5)$, we see from Eq.~(\ref{eq:threxp}) 
that the expansion parameter is $a/L$. Because of this, for a fixed range of $L$,
we expect the expansion to break down as $|a|$ increases. This is borne out by
the results shown in Fig.~\ref{fig:EvsLVaryA}, where only for $|a|\lesssim 1$ does the
threshold expansion---shown only through $\cO(1/L^5)$ in the plots---provide a
good description over most of the range of $L$.

Three-particle interactions enter Eq.~(\ref{eq:threxp}) only at $\cO(1/L^6)$,
through the quantity $\Mthr$, which is a particular definition of the three-particle
divergence-free scattering amplitude at threshold, and is discussed in Sec~\ref{sec:KtoMthr}
and Appendix \ref{app:impint}.
As noted earlier, the appearance only at high order implies
that the spectrum is only sensitive to three-particle interactions at smaller values of $L$, 
which is the region where simulations are done.
But in this small $L$ region, the finite-order threshold expansion might not apply, and one must then use the
full quantization condition.
By contrast, in this section we are aiming to test our numerical methods by working in
a regime where the threshold expansion does hold, namely small $|a|$ and large $L$.
Specifically, we consider a single value of the scattering length, $a=0.41315$, 
and determine the threshold energy to very high accuracy for the range $L=5-60$,
and with $1/\Kiso=0.04-0.16$.
By doing so we are able to extract a value for $\Mthr$---which is the only undetermined
parameter in Eq.~(\ref{eq:threxp}). Our results for $\Mthr$ as a function of
$\Kiso$ can then be checked against the predictions from the infinite-volume
integral equations, as will be discussed in Sec.~\ref{sec:KtoMthr} below.

\begin{figure}[tbh]
\begin{center}
%\vskip -.2truein
%\includegraphics[scale=0.6]{thra04.pdf}
%\includegraphics[width=\textwidth]{EthrvsL_KSvsHSSLabeled.pdf}
\includegraphics[width=\textwidth]{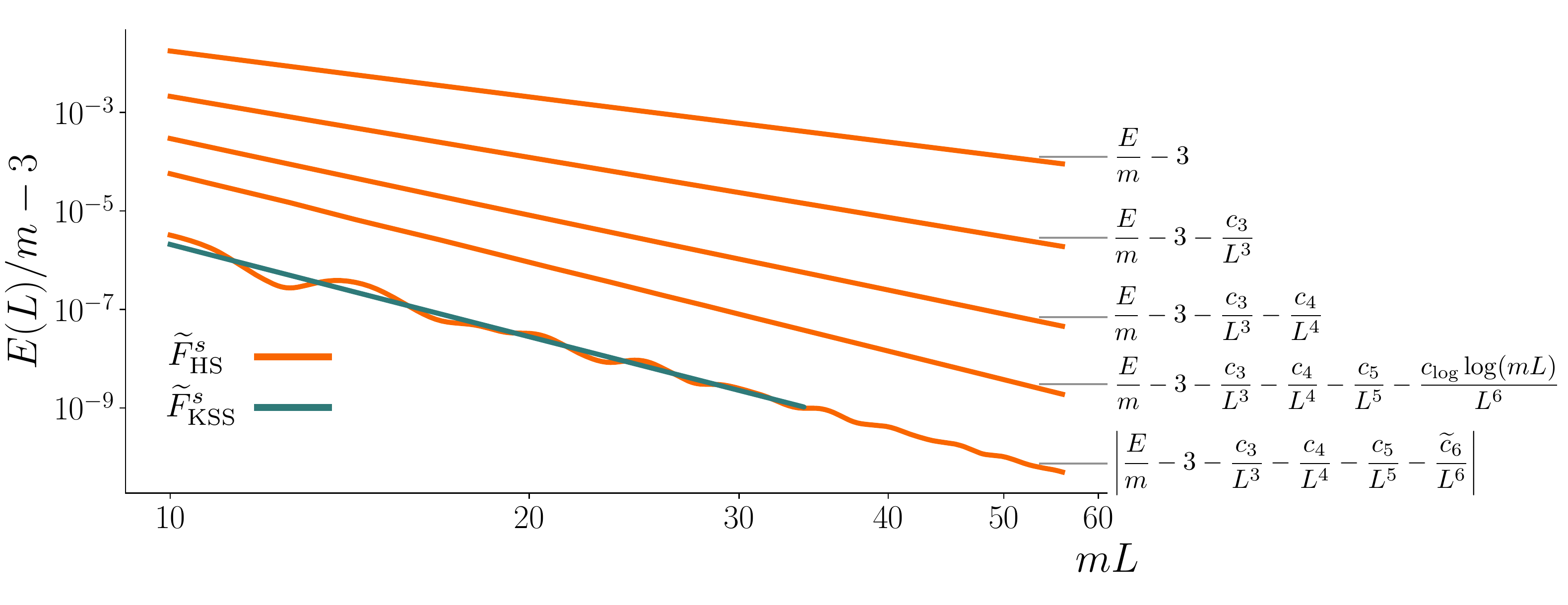}
%\vskip -4.2truein
\caption{
Log-log plot of $E(L)/m-3$ vs.~$mL$ (top curve) determined from the quantization condition for $ma=0.41315$ and $m^2 \Kiso=10$, together with various subtracted curves as labeled. 
The results are indistinguishable for the two definitions of $\widetilde F_s$, with the exception of the lowest (maximally-subtracted) curve, where the $H$-function regulator is shown in orange and that based on Ref.~\cite{Kim:2005gf} in blue. The oscillations in the former are discussed in the text.
}
\label{fig:thra04}
\end{center}
\end{figure}

In Fig.~\ref{fig:thra04} we show that the numerical results from the quantization condition are very well described by the threshold expansion for our choice of scattering parameters. The top curve shows the results from the quantization condition for $E(L)-3$. 
Here we suppress the comparison to $c_3/L^3+c_4/L^4+c_5/L^5$ as the curves are 
indistinguishable at this scale 
(indeed, $c_3/L^3+c_4/L^4$ is already indistinguishable from the top curve). 
The plot also shows the residuals
as successively more terms are subtracted from the threshold expansion, as labeled.
We see nice convergence for $L \gtrsim 10$, with each successive term 
improving the agreement, and the residuals decreasing in the expected way with $L$. 
Note that we subtract the log-dependent piece of $\widetilde c_6$ together with $c_5$ in the second-to-last residual, as these terms are of similar numerical magnitude.
We stress that we must solve the quantization condition with a numerical
accuracy of better than 1 part in $10^8$ in order to pick out the maximally-subtracted
result. This turns out to be straightforward.

The maximally-subtracted residual shows oscillatory behavior.
To investigate this, we have repeated the calculation replacing the
sum-integral difference regulated using $H$-functions, $\widetilde F^s_{\rm HS}$,
with that regulated following Ref.~\cite{Kim:2005gf}, $\widetilde F^s_{\rm KSS}$.
The residues are indistinguishable for all but the lowest curve,
in which we find that the results obtained
using $\widetilde F^s_{\rm KSS}$ do not oscillate.
Since the difference between the two choices of $\widetilde F^s$ is exponentially suppressed,
we conclude that the oscillations represent
a class of neglected exponentially-suppressed finite-volume effects.
They are visible here presumably because we are investigating tiny contributions to the energy.
Other examples of such effects will be seen below.

\begin{figure}[tbh]
\begin{center}
%\vskip -.2truein
%\includegraphics[scale=0.6]{res6a04.pdf}
\includegraphics[width=0.75\textwidth]{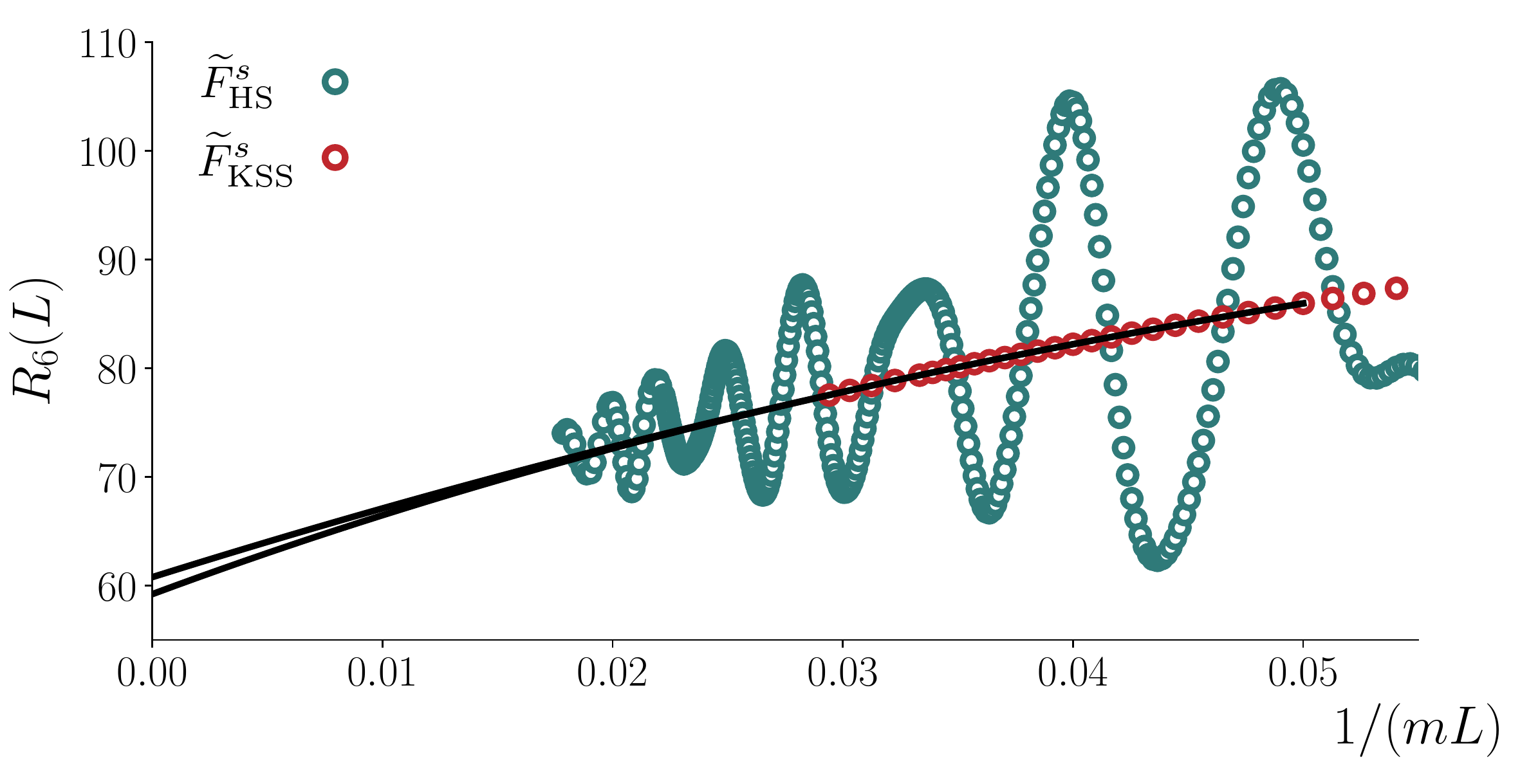}
%
%\vskip -4.2truein
\caption{
Plot of $R_6(L)$ [defined in Eq.~(\ref{eq:R6})] versus $1/(mL)$ for $ma=0.41315$ and
$m^2\Kiso=10$. The oscillating (blue) points use $\widetilde F^s_{\rm HS}$, while the
smooth (red) points use $\widetilde F^s_{\rm KSS}$.
The solid curves show quadratic and cubic fits in $1/(mL)$ to the 
$\widetilde F^s_{\rm KSS}$ data up to $1/(mL)=0.05$. We take the average of these curves at $1/(mL)=0$ as the central value for the infinite-volume limit, and half the difference as the uncertainty}
\label{fig:res6a04}
\end{center}
\end{figure}

As noted above, we can determine $\Mthr$ from the maximally-subtracted results.
To do so, we scale up the residual by $L^6$ and define
\begin{equation}
 R_6(L) \equiv 
- L^6\left\{ E(L) - 3 - \frac{c_3}{L^3} - \frac{c_4}{L^4} - \frac{c_5}{L^5} - \frac{\wt c_6}{L^6} \right\}
= \frac{\Mthr}{48} + \cO(1/L)
\,.
%L^6  R_6 \equiv 
%L^6\left\{\frac{c_3}{L^3} + \frac{c_4}{L^4} + \frac{c_5}{L^5} + \frac{\wt c_6}{L^6}
%- (3-E)\right\}
%= \frac{\Mthr}{48} + \cO(1/L)
%\,,
\label{eq:R6}
\end{equation}
This quantity is shown in Fig.~\ref{fig:res6a04} as a function of $1/L$ for $L\gtrsim 20$.
Here we again show the results using the two regulators for $\wt F^s$. 
The oscillations with $\widetilde F^s_{\text{HS}}$ are more pronounced 
with the new scale, and it is easier to use the $\widetilde F^s_{\text{KSS}}$ 
results to extrapolate to the infinite-volume limit.
Averaging quadratic and cubic fits in $1/L$ to the latter yields $\Mthr/48=60.0 \pm 0.8$,
with the uncertainty determined by half the difference between the two fits. 

\begin{figure}[tbh]
\begin{center}
%\vskip -.2truein
%\includegraphics[scale=0.6]{extrapderiv}
\includegraphics[width=0.75\textwidth]{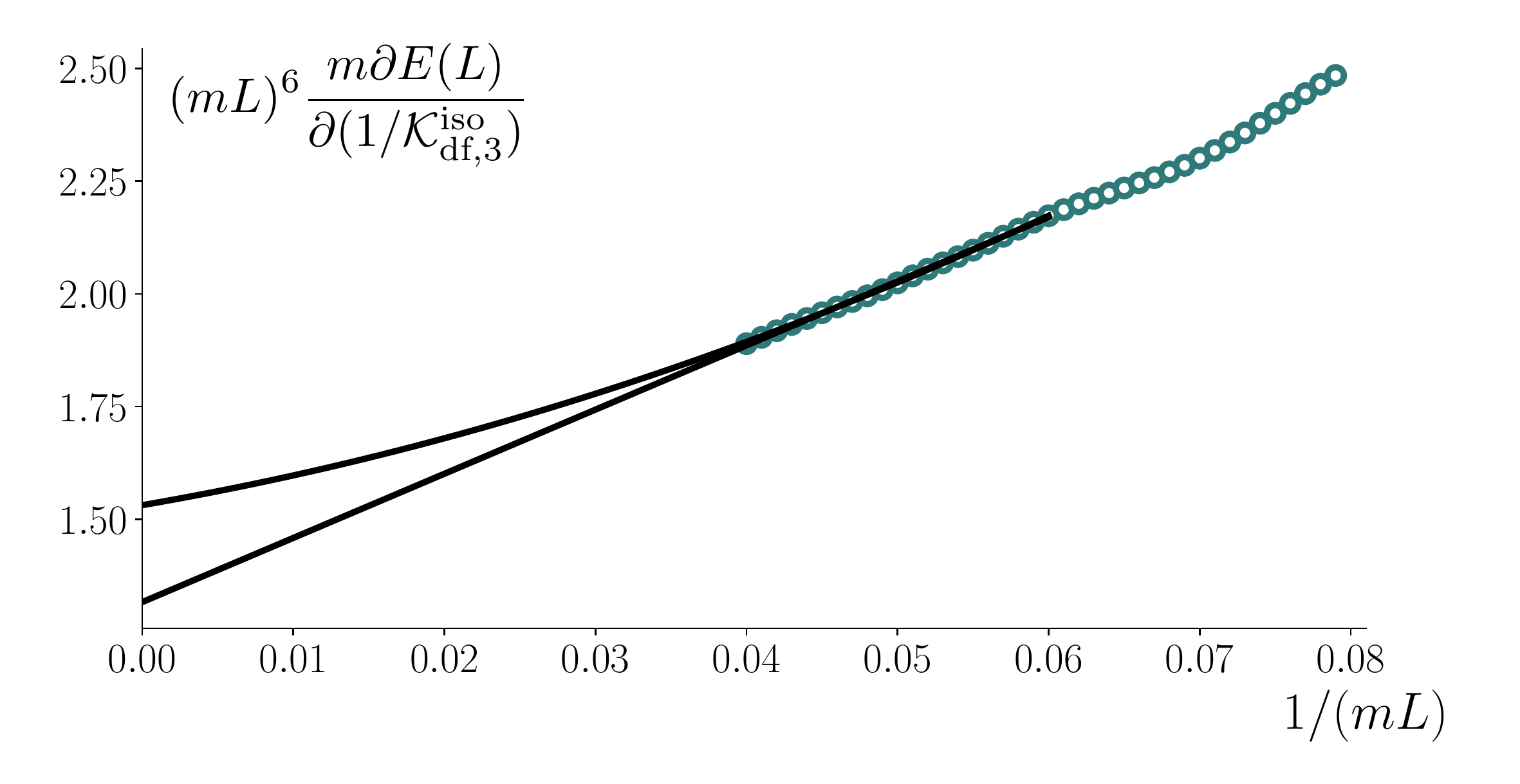}
%\vskip -4.2truein
\caption{
Extrapolation in $1/(mL)$ of the left-hand side of Eq.~(\ref{eq:derivmthr}) 
evaluated at $1/(m^2\Kiso)=0.1$ and $ma=0.41315$.
Linear and quadratic fits are done to the region of points indicated by the curves. 
{We stress that this data was generated using $\widetilde F^s_{\text{HS}}$,
but in this case there are only weak oscillations, unlike in Fig.~\ref{fig:res6a04}.}
}
\label{fig:extrapderiv}
\end{center}
\end{figure}

\begin{figure}[tbh]
\begin{center}
%\vskip -.2truein
%\includegraphics[scale=0.6]{derivmthrf}
\includegraphics[width=0.75\textwidth]{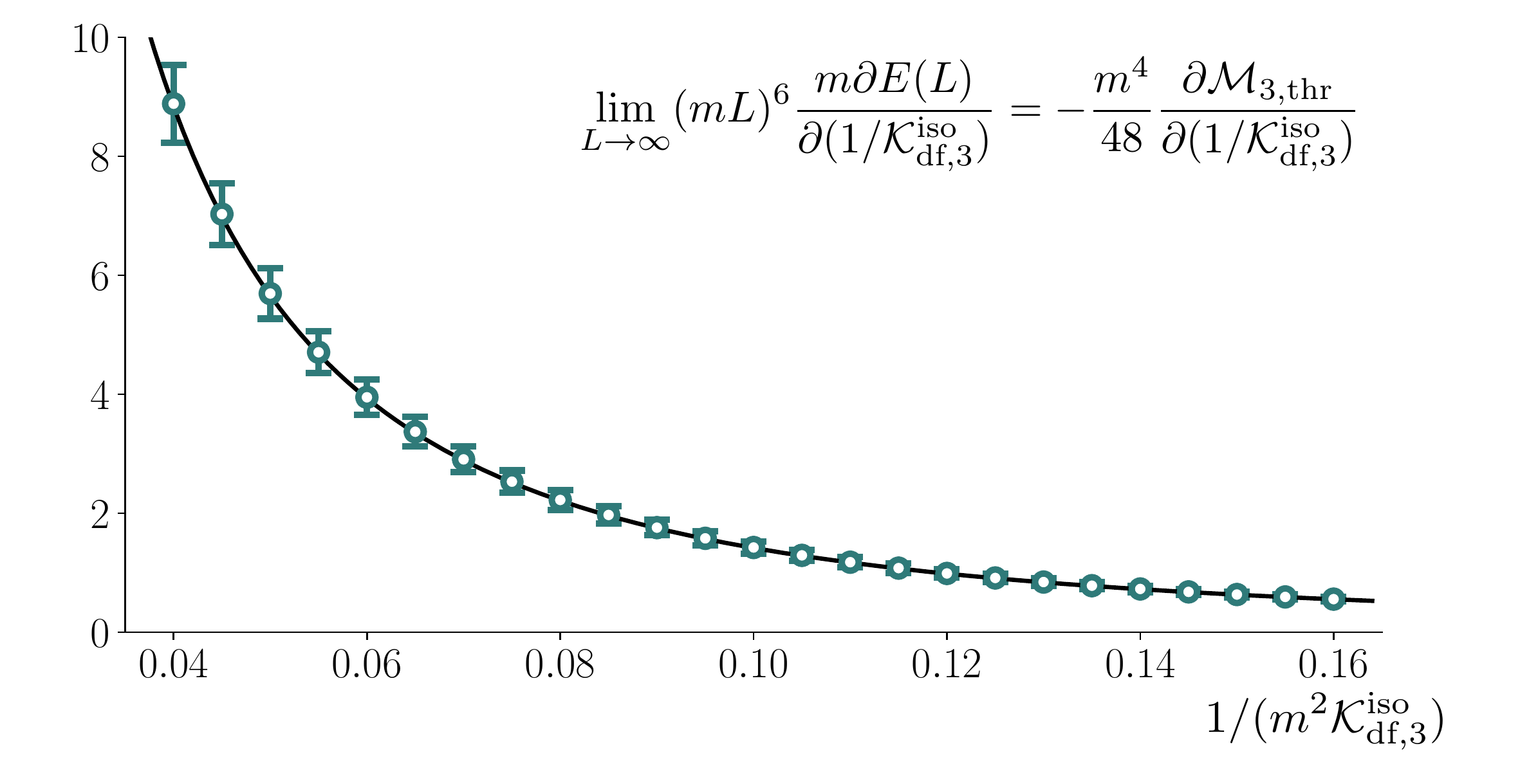}
%\vskip -4.2truein
\caption{
Dependence of $-(m^4/48) {\partial \Mthr}/{\partial (1/\Kiso)}$ vs.~$1/(m^2\Kiso)$ for $ma=0.41315$.
The points are obtained from the threshold energy, using
extrapolations such as that in Fig.~\ref{fig:extrapderiv},
while the solid curve
is obtained from solving the integral equation relating $\Mthr$ and $\Kiso$.
}
\label{fig:derivmthrf}
\end{center}
\end{figure}

We close this subsection by considering one additional infinite-volume quantity that 
can be extracted from the threshold energy. 
With little additional effort we can determine the dependence of the extracted
$\Mthr$ on $\Kiso$, using\footnote{%
We take the derivative with respect to $1/\Kiso$
because this, rather than $\Kiso$ itself, is the more natural quantity entering the quantization condition in the form we use.}
\begin{equation}
L^6 \frac{ \partial E(L)}{ \partial (1/\Kiso)}\bigg|_{a,L} = 
- \frac{1}{48} \frac{\partial \Mthr}{\partial (1/\Kiso)} \bigg|_{a,E=3}
+ \cO\left(\frac1{L}\right)\,.
\label{eq:derivmthr}
\end{equation}
We determine the derivative numerically by varying $\Kiso$ close to $10$.\footnote{%
Given the weak dependence of $E$ on $\Mthr$ we need to vary $E$ over a very small range.
For example, for $L=20$, the range $E=3.002067695-3.002067697$ leads to
a variation in $\Kiso$ from $\approx 6-13$ when $a=0.41315$.}
The extrapolation to $L=\infty$ is done either linearly or quadratically in $1/L$.
An example is shown in Fig.~\ref{fig:extrapderiv}.
{Comparing to the results for $R_6(L)$,
we see that the derivative removes
much of the oscillatory volume dependence,
in addition to the first three orders in $1/L$.}
We show the resulting $\Kiso$ dependence of the extrapolated
derivative in Fig.~\ref{fig:derivmthrf}. 
We take the average of linear and quadratic extrapolations as the central value and half the difference as the uncertainty. 
The solid line shows the infinite-volume prediction found by solving the integral 
equation relating $\Mthr$ to $\Kdf$, discussed in Sec.~\ref{sec:KtoMthr} below.
We stress that this is not a fit to the data, but rather the result of an independent
calculation. The agreement between the two results provides a strong check of
our numerical implementation of the quantization condition, as well as of the
analytic derivation of the threshold expansion in Ref.~\cite{\HSTH}.

\section{Relating $\Kiso$ to the scattering amplitude \label{sec:thresh}}

{

As explained in Sec.~\ref{sec:formalism}, to obtain physical infinite-volume quantities
given knowledge of $\Kiso$ requires solving an integral equation and performing several
integrals.
In this section we show how this can be done by straightforward
extensions of the numerical methods used to solve the quantization condition,
as long as we work below or at threshold.
We divide this section into three parts. 
In the first we show results for the quantities needed to relate $\Kiso$ to $\Mdf$
below threshold.
In the second, we show how, in the case of a three-particle bound state, we can
determine a quantity related to the infinite-volume Bethe-Salpeter amplitude.
This quantity can then be compared to the predictions of NRQM.
Finally, we work directly at threshold and 
 calculate the relation between $\Kiso$ and the
quantity that enters into the threshold expansion, $\Mthr$. 
}

\subsection{Relating $\Kiso$ to $\Mdf$ below threshold}
\label{sec:KtoMsubthr}

The relationship between $\Kiso$ and $\Mdf$ is given in Eq.~(\ref{eq:M3dKiso}), 
which we reproduce here for clarity, making use of the results that
$\cL(\vec k)=\cR(\vec k)$ and that $\cL$ depends only on the magnitude of $k$
in the isotropic approximation,
\begin{align}
\Mdf (\vec k, \hat a^*;\vec p, \hat a'^*) &= 
\cS\left\{\frac{\cL(k) \cL(p)}{1/\Kiso + F_3^\infty} \right\}\,.
\label{eq:M3dKiso2}
\end{align}
In this subsection we illustrate how to
calculate the quantities on the right-hand side of this equation when
working below threshold.
The methods for doing so are explained in Appendix~\ref{app:impint}.
The infinite-volume quantities
$F_3^\infty$ and $\cL(k)$ can be obtained simply by taking the 
$ L\to\infty$ limit of appropriate finite-volume quantities.
In the case of $F_3^\infty$ one choice of finite-volume quantity is simply $F_3^\iso$.

\begin{figure}[tbh]
\begin{center}
\includegraphics[width=0.75\textwidth]{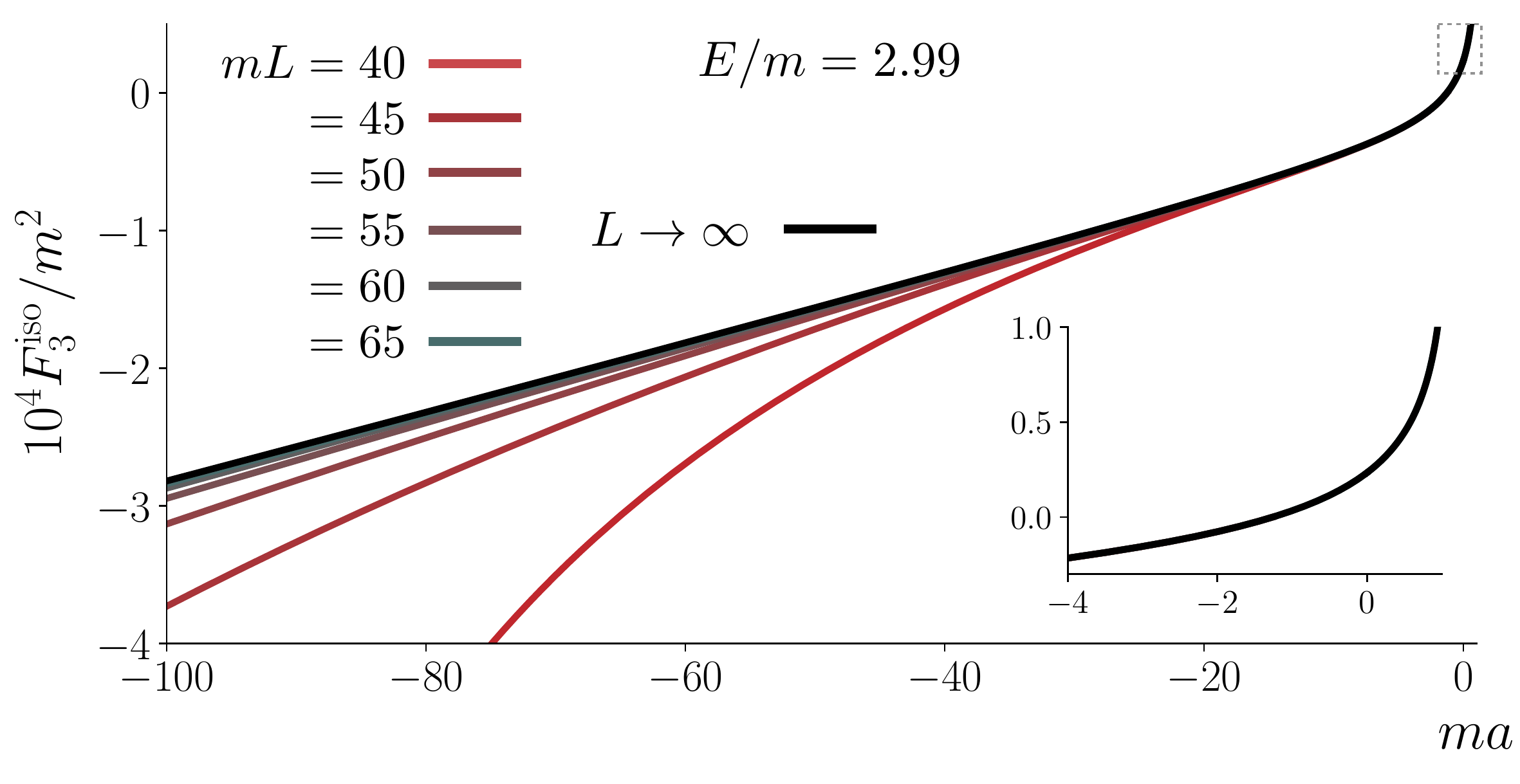}
\caption{
$F_3^\iso/m^2$ vs.~$ma$ for $E=2.99m$ and $mL=40-65$, 
together with an extrapolation to $L \to \infty$ using $mL=50-65$.
The inset shows the small $ma$ region, in which $F_3^\iso$ changes sign.
}
\label{fig:F3E299}
\end{center}
\end{figure}

\begin{figure}[tbh]
\begin{center}
 \includegraphics[width=0.75\textwidth]{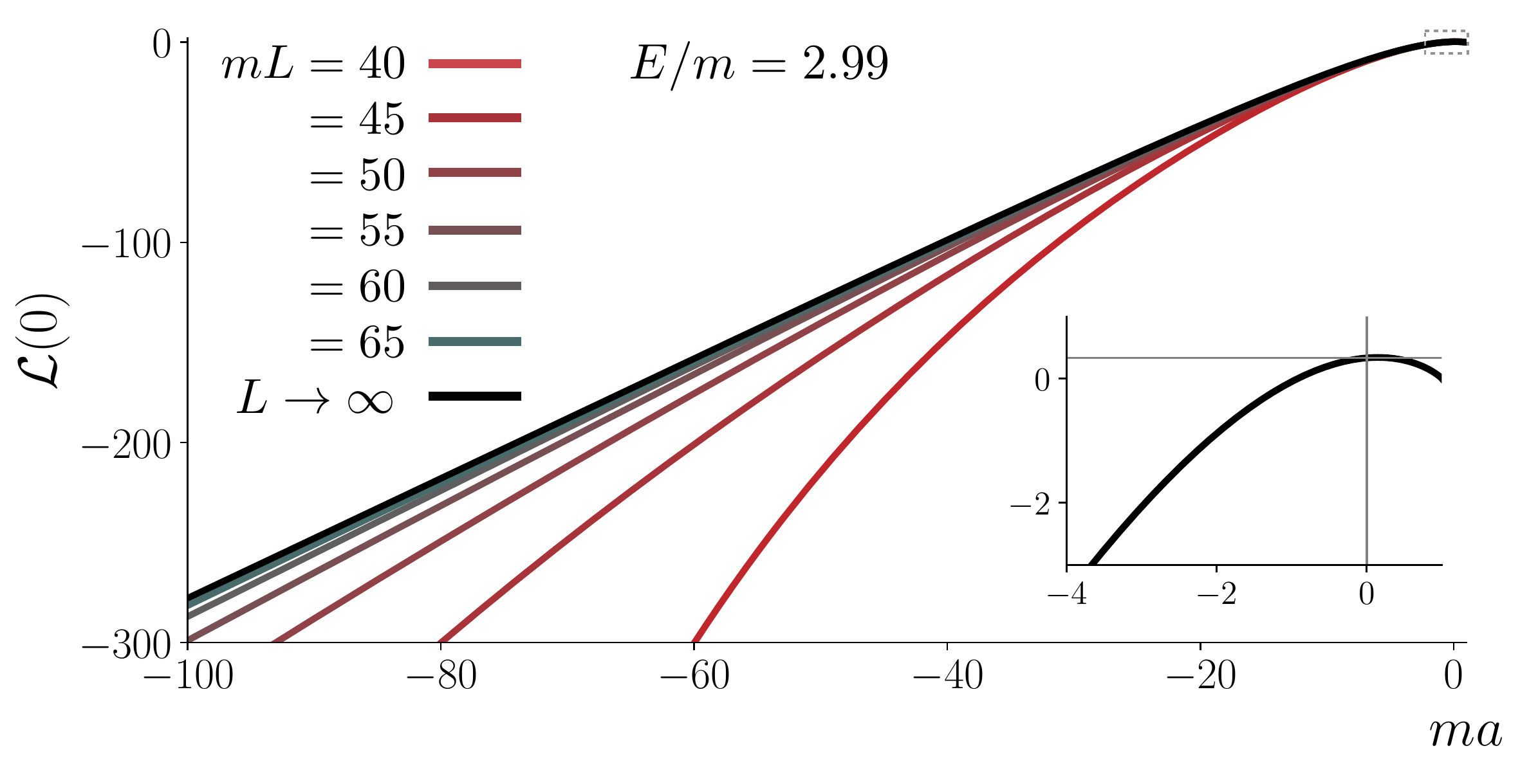}
\caption{
${\cal L}(\vec k = 0)$ vs.~$ma$ for $E=2.99m$ and $mL=40-65$, 
together with an extrapolation to $L \to \infty$ using $mL=50-65$.
Here ${\cal L}(\vec k)$ for finite $L$ is given by Eq.~(\ref{eq:calLinflim}).
The inset shows the small $ma$ region, within which $\cL(0)$ changes sign.
Note that $\cL(0)=1/3$ when $m a=0$.
}
\label{fig:LE299}
\end{center}
\end{figure}

In Figs.~\ref{fig:F3E299} and \ref{fig:LE299} 
we show the approach to the $L=\infty$ limit 
for $F_3^\iso$ and $\cL(0)$, respectively, taking $E=2.99$ as an example.
For fixed $a$, the approach to the limit is exponential, 
allowing a controlled extrapolation to $L=\infty$,
although larger values of $L$ are needed as $|a|$ increases.

Figure~\ref{fig:F3E299} illustrates why, generically,
there are bound states for a range of values of $\Kiso$. We recall that, for any finite $L$,
there is a solution to the quantization condition if $F_3^\iso=-1/\Kiso$.
Since $F_3^\iso$ approaches a limiting function of $a$ as $L\to\infty$, 
which we observe to be monotonically increasing,
there will be a bound state with $E=2.99$ at some value of $a$ for all values of 
$\Kiso$ in the range $-1/F_3^\iso(a=1) < \Kiso < -1/F_3^\iso(a=-\infty)$.
Since the limiting function is negative for almost all values of $a$, most bound states
occur with positive values of $\Kiso$. 
One example (for a different value of $E$) is the bound state
discussed in Sec.~\ref{sec:bound}.

\begin{figure}[tbh]
\begin{center}
\includegraphics[width=0.75\textwidth]{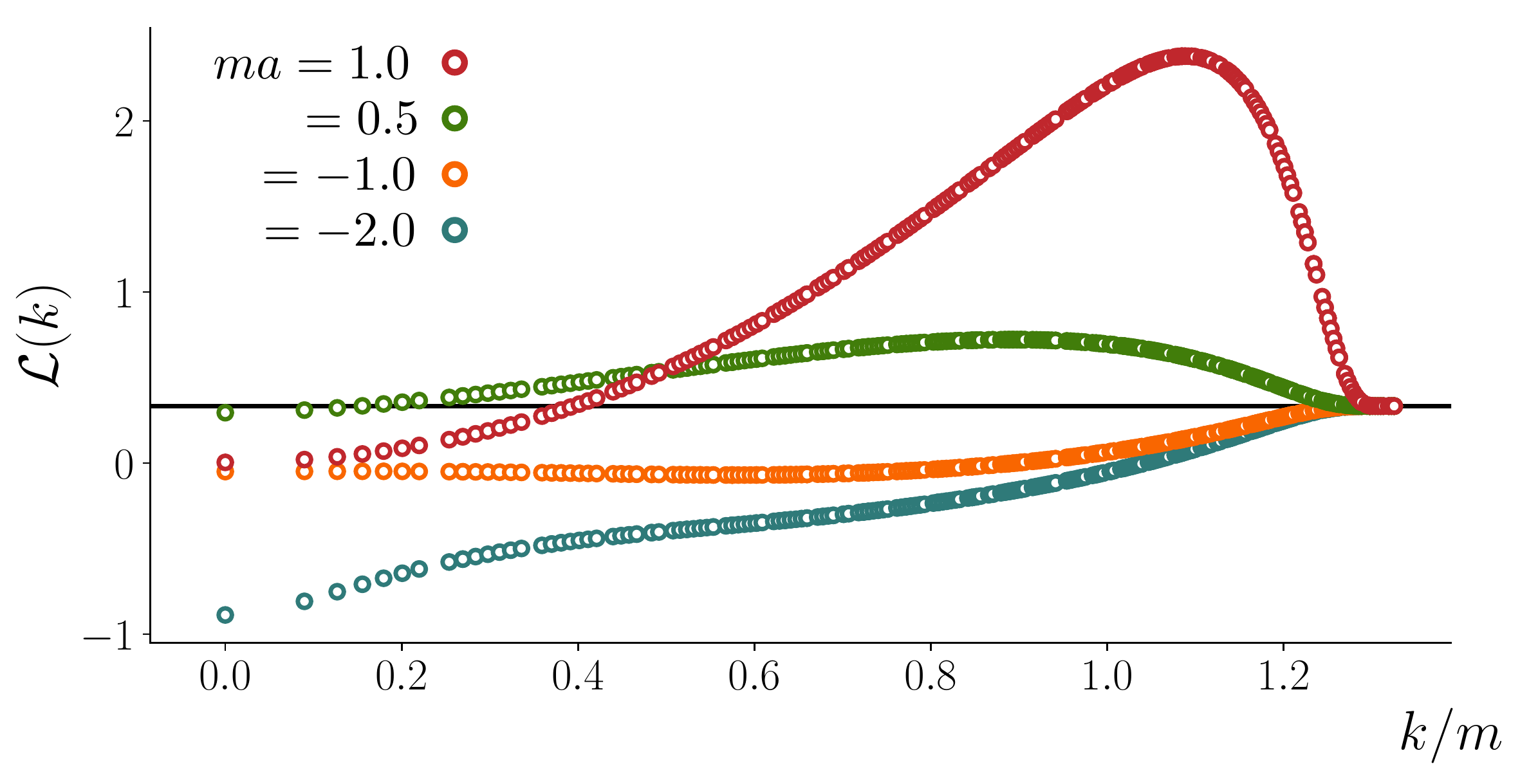}
\caption{
$\cL(k)$ versus $k/m$ for choices of $ma$ shown in the legend. Results using
either choice of finite-volume quantity, Eq.~(\ref{eq:calLinf}) or (\ref{eq:calLinflim}),  
and using any choice of $mL \ge 50$, lie on a common curve.
Here we show the results using Eq.~(\ref{eq:calLinflim}) and $mL=70$.
Note that, if $a=0$, $\cL(k)=1/3$ independent of $k$. 
For sufficiently large $k$, $\cL(k)=1/3$ for all $a$, 
since the cutoff functions vanish and remove the correction term.
}
\label{fig:ellk}
\end{center}
\end{figure}

Figure~\ref{fig:ellk} shows examples of the $k$-dependence of $\cL(k)$ for various
choices of $a$. This quantity describes the effect of multiple two-to-two scattering
with the scattering pair changing each time to include the spectator of the previous event.
As $k$ increases the scattered pair lies increasingly far below threshold.
For a bound state, $\cL(k)$ is related to the Bethe-Salpeter amplitude,
as discussed in the following subsection.

The results for $F_3^\infty$ and $\cL(k)$ can be combined to determine 
results for $\Mdf$, using Eq.~(\ref{eq:M3dKiso2}).
We choose not to quote results here since
the symmetrization that is needed is complicated, and the results produced are not
transparent. We will, however, quote the corresponding results below when working
at threshold.

\subsection{Determining the wavefunction of the bound state}
\label{sec:BS}

A specific application of the subthreshold relation between $\Kiso$ and $\Mdf$
is provided by the bound state studied in Sec.~\ref{sec:bound}. 
For the fixed values of $\Kiso=2500$ and $a=-10^4$, 
one can calculate $F_3^\infty$ and identify the infinite-volume bound state pole in $\Mdf$,
as described in the previous subsection. 
Since this is equivalent to solving the quantization condition $\Kiso = - 1/F_3^{\text{iso}}$ 
for asymptotically large volumes, one finds the same result for 
the infinite-volume bound-state energy as from the fit in Sec.~\ref{sec:bound},
namely $E_B = 2.98858$ (corresponding to $\kappa = 0.106844$). 

\begin{figure}[tbh]
\begin{center}
\includegraphics[width=0.8\textwidth]{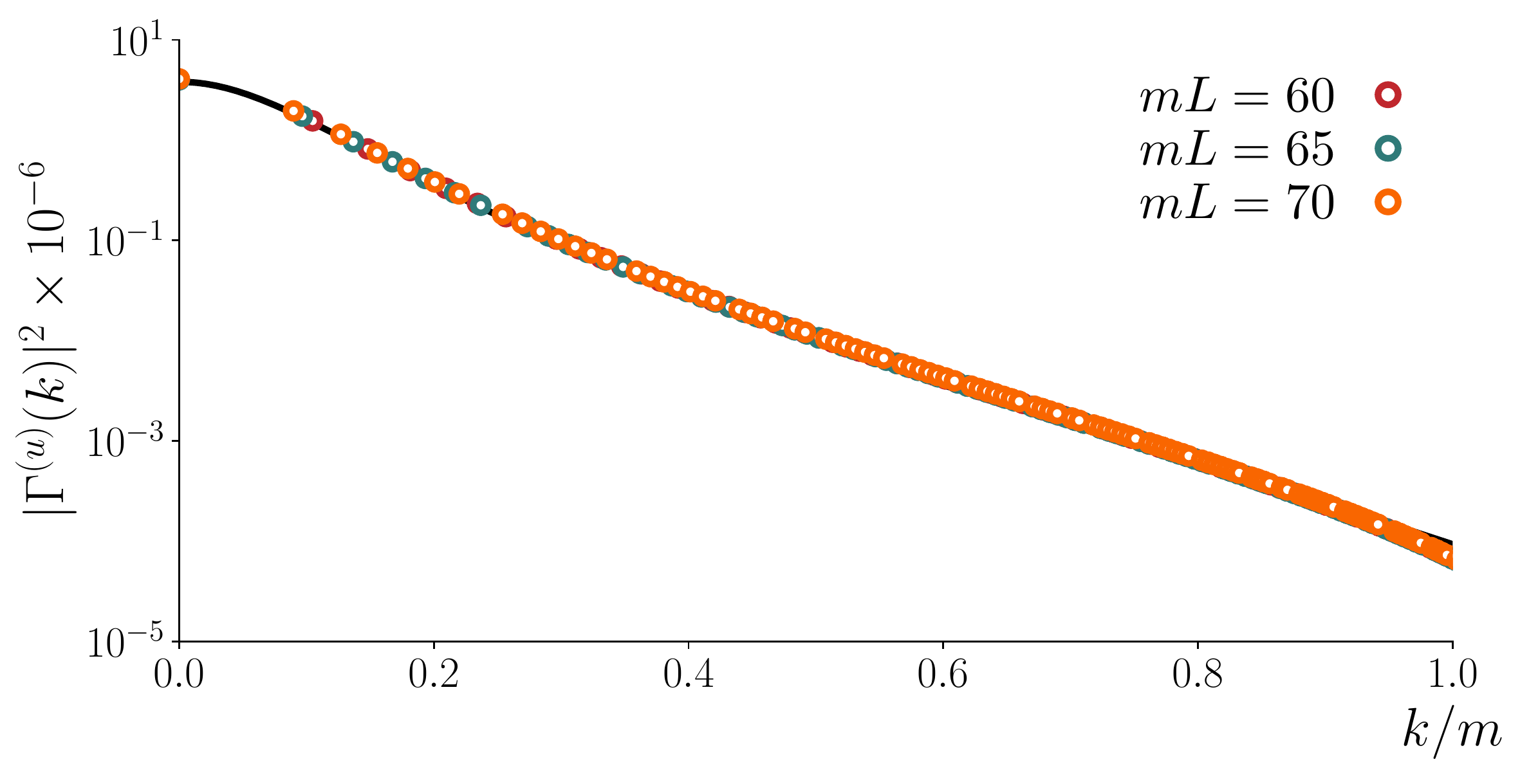}
\caption{
Momentum dependence of the {magnitude squared of the} bound-state residue function. 
The points are predictions following from Eqs.~(\ref{eq:residue2}) and (\ref{eq:residue3}),
as described in the text. Different values of $L$ lead to consistent results, indicating that
we have reached the infinite-volume limit.
The curve shows the prediction of Eq.~(\ref{eq:residuepred}), with the value
$|A|^2=0.948$ found in Sec.~\ref{sec:bound}.
\label{fig:BoundGammavsk} }
\end{center}
\end{figure}

The residues of the pole in $\Mdf$ contain information about the Bethe-Salpeter 
amplitudes of the bound state. Specifically, as discussed in Ref.~\cite{\HSBS},
the unsymmetrized version of $\Mdf$ takes the following factorized form near the bound state
\begin{equation}
\mathcal M^{(u,u)}_{\text{df},3}(k,p) \sim  - 
\frac{  \Gamma^{(u) }(k) \Gamma^{(u) }(p)^* }{E^2 - E_B^2} \,.
\end{equation}
This assumes that pairwise scattering occurs only in the $s$-wave,
as is the case in the isotropic approximation.
The quantity $\Gamma^{(u)}(k)$ is related to the Bethe-Salpeter amplitude by
amputating and going on shell, as explained in detail in Appendix B of Ref.~\cite{\HSBS}.
We call $\Gamma^{(u)}(k)$ the residue function.
Combining this expression with Eq.~(\ref{eq:M3dKiso2}) we find
that $\Gamma^{(u)}(k)$ is proportional to $\cL(k)$,
\begin{equation}
 \vert \Gamma^{(u) }(k) \vert^2   =  
 \lim_{E \to E_B} (E_B^2 - E^2) \frac{\cL( k)^2}{1/\Kiso(E) + F_3^\infty(E)} \,.
\label{eq:residuesq}
\end{equation}
In our approach both $F_3^\infty(E)$ and $\cL(k)$ are determined 
by taking infinite-volume limits of appropriate finite-volume quantities. 
For the purposes of extracting $ \vert \Gamma^{(u) }(k) \vert^2$ it turns out to be 
convenient to define a finite-volume version as
\begin{equation}
 \vert \Gamma^{(u) }(k) \vert^2(L)   =  
 \lim_{E \to E_B(L)} (E_B^2(L) - E^2) \frac{\cL_L(E, k, L)^2}{1/\Kiso(E) + \Fiso(E,L)} \,,
 \label{eq:residue2}
\end{equation}
where $\cL_L(E, k, L)$ is defined as the argument of the limit in Eq.~(\ref{eq:calLinflim}). 
Using this quantity, the infinite-volume limit,
\begin{equation}
 \vert \Gamma^{(u) }(k) \vert^2   = \lim_{L \to \infty}  \vert \Gamma^{(u) }(k) \vert^2(L)  \,,
 \label{eq:residue3}
\end{equation}
is approached more rapidly.
Figure~\ref{fig:BoundGammavsk} shows numerical results for
$\vert\Gamma^{(u)}(k)\vert^2(L)$, calculated by setting $E=E_B(L)+\delta E$ 
(with $\delta E=-0.001$) and using $mL=60,65,70$. The results fall on a common
 curve giving confidence that we have reached the infinite-volume limit. 

In Ref.~\cite{\HSBS} we showed that, in NRQM in the unitary limit, 
the residue function is given by\footnote{%
It is interesting to note that the leading finite-volume dependence of the bound state energy,
given in Eq.~(\ref{eq:FVBS}), is obtained using the leading term in the expansion of the
result presented here
for $\Gamma^{(u)}(k)$ about the singularity at $k^2=-\kappa^2$. 
This leading term is given in Eq.~(100) of Ref.~\cite{\HSBS}.
When evaluated on the real axis, however,
it differs substantially from the full result.
Thus it is essential to use the full form given here when studying the function for real $k$}.
\begin{equation}
 \vert \Gamma^{(u) }(k)_{\text{NR}}\vert^2 = \vert c \vert \vert A \vert^2  \frac{256 \pi ^{5/2}}{3^{1/4}}   \frac{ m^2 \kappa^2 }{ k^2  (\kappa ^2+{3 k^2}/{4} )}
  \frac{\sin ^2 \Big(s_0 \sinh^{-1} \frac{\sqrt{3} k}{2 \kappa } \Big) }{\sinh^2\frac{\pi  {s_0}}{2} }  \,,
 \label{eq:residuepred}
\end{equation}
with $s_0=1.00624$ and $|c|=96.351$, and $|A|$ the quantity
entering into Eq.~(\ref{eq:FVBS}).
This prediction is also plotted in Fig.~\ref{fig:BoundGammavsk},
and is in excellent agreement with our numerical results.
We stress that this curve is a parameter-free prediction and not a fit.
However, we do expect there to be relativistic corrections to the relationship between
$\Gamma^{(u)}(k)$ and $\Gamma^{(u)}(k)_{\rm NR}$.
These should vary in magnitude between of
$\cO(\kappa^2/m^2)=\cO(1\%)$ at $k=0$ to of $\cO(k/m)=\cO(1)$ for $k\approx m$.
These expectations are consistent with the small differences we find.

What do learn from this agreement?
The derivation of Eq.~(\ref{eq:residuepred}) in Ref.~\cite{\HSBS} does not
use the quantization condition in any way. 
Instead, it relies only on the definition of the relativistic scattering amplitude and 
the standard NRQM determination of the bound-state wave function.
Thus the agreement is not a consistency check, but rather shows that the
relation~(\ref{eq:M3dKiso2}) reproduces the physics leading to 
the Efimov bound-state solution of the NRQM problem. 
This is also true for the predicted volume dependence of the bound-state energy,
discussed in Sec.~\ref{sec:bound}, but here the test is even more stringent because
we are predicting a function and not just a number.

Finally, we note that the curves in Fig.~\ref{fig:ellk} are proportional to
the residue functions for bound states that are not in the unitary regime.
This is because, for all values of $a < 1$, one can tune $\Kiso$ to give 
a bound state at $E=2.99$, and then use Eq.~(\ref{eq:residuesq}).
Since the $k$ dependence comes only from $\cL(k)$, 
it follows that $|\Gamma^{(u)}(k)| \propto |\cL(k)|$.
We observe that, away from the unitary regime,
the dependence on $k$ varies substantially with $a$.
It would be interesting to compare these results to predictions from NRQM.

\subsection{Relating $\Kiso$ at threshold to $\cM_{3,\df,\thr}$ and $\Mthr$}
\label{sec:KtoMthr}

As discussed above, $\Mdf$ is finite for all energies and choices of momenta
(aside from bound-state poles). 
In particular it is finite at threshold, and we denote its value there by $\Mdfthr$.
This divergence-free scattering amplitude is defined by subtracting an infinite series 
of terms from the usual three-to-three scattering amplitude, $\cM_3$. 
An alternate definition of a finite, three-particle threshold amplitude was 
introduced in Ref.~\cite{\HSTH}, based on subtracting from $\cM_3$ only those
parts of $\cD^{(u,u)}$ that contain IR divergences.
 It is this new quantity, called $\Mthr$,
that appears in the threshold expansion, Eq.~(\ref{eq:threxp}) above. 

In Sec.~\ref{sec:threxp} above, we studied the threshold state predicted by the quantization condition for $a=0.41315$ and $\Kdf=10$ and found that the volume dependence is very well described by the threshold expansion with $\Mthr/48=60.0 \pm 0.8$. 
In this section we aim to test this result by directly applying the relation 
between $\Kiso$ and $\Mdfthr$ as well as that between 
 $\Mdfthr$ and $\Mthr$~\cite{\HSTH}. 

We begin by solving the integral equations relating $\Kiso$ to $\Mdfthr$. 
At threshold, the general relationship of Eq.~(\ref{eq:M3dKiso2}) simplifies to
\begin{align}
\Mdfthr = 
 \frac{9 \cL(0)^2}{1/\Kiso + F_3^\infty}   \,,
\label{eq:MdftoKdfthr}
\end{align}
with the factor of $9$ arising from symmetrization.
Thus we need only to determine $\cL(0)$ and $F_3^\infty$ at $E=3$, for the
chosen value of $a$.
 This is slightly more complicated than the subthreshold determinations discussed 
 above because the finite-volume analogs of $\cL$ and $F_3^\infty$ both diverge for $E=3$. 
 We have two methods to circumvent this issue. 
 One option is to take the $L \to \infty$ limit for a set of sub-threshold values of $E$
 (using the method described in previous subsections)
 and then extrapolate $E \to 3$. 
 An alternative, direct approach is to define modified versions of the finite-volume 
 objects in which the singularity at $E=3$ is removed. 
 As explained in Ref.~\cite{\HSTH}, this removal does not affect the $L \to \infty$ limit.
The direct approach has the advantage that only one limit need be considered.
We have confirmed that the two methods give consistent results and in this subsection 
only show results using the direct approach. 
The details of its numerical implementation are summarized in Appendix~\ref{app:impintthr}.

\begin{figure}[tbh]
\begin{center}
%\vskip -.2truein
%\includegraphics[scale=0.6]{f3infa4}
\includegraphics[width=0.7\textwidth]{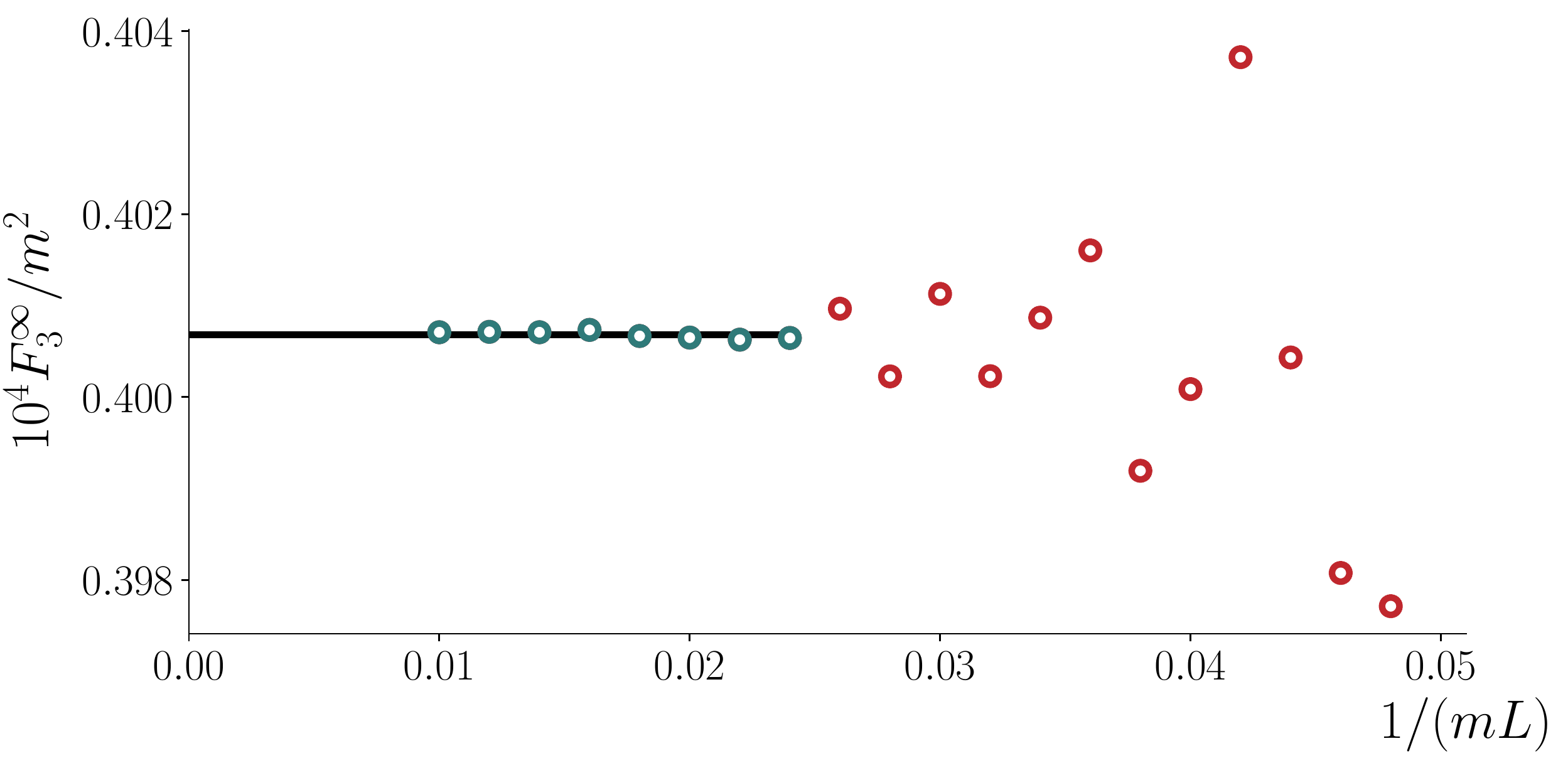}
%\vskip -4.2truein
\caption{
$F_3^\infty/m^2$ vs. $1/(mL)$ for $ma=0.41315$ at threshold. 
The line indicates a fit of the points shown in blue to a constant,
which we use as our estimate of the $L \to \infty$ value.
}
\label{fig:f3infa4}
\end{center}
\end{figure}
\begin{figure}[tbh]
\begin{center}
%\vskip -.2truein
%\includegraphics[scale=0.6]{ell0a4}
\includegraphics[width=0.75\textwidth]{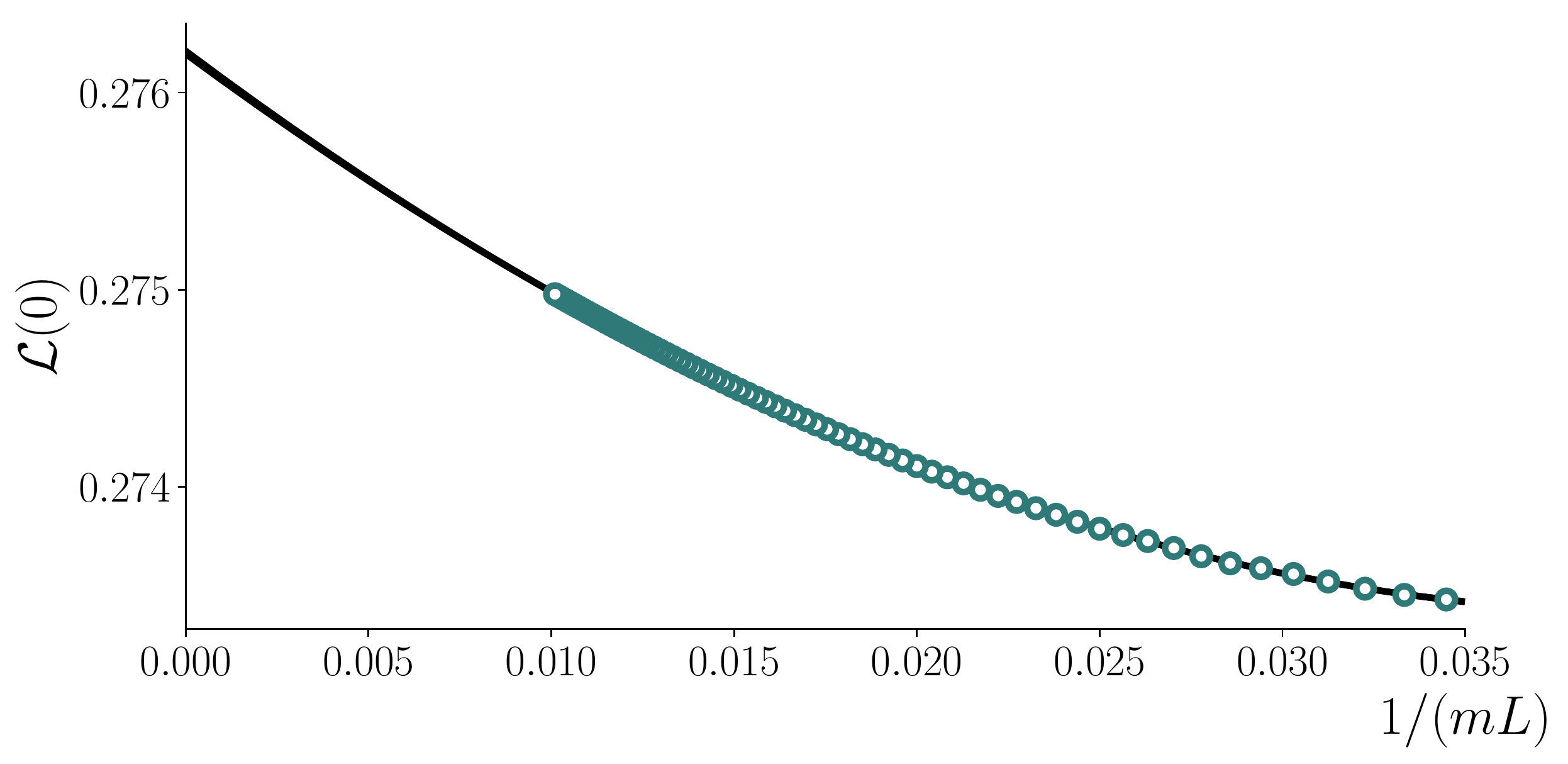}
%\vskip -4.2truein
\caption{
$\cL( 0)$ vs. $1/(mL)$ for $ma=0.41315$ at threshold.  
Quadratic and cubic fits to the entire set give very consistent results at $L=\infty$. (Both curves are plotted but are indistinguishable aside from a thickening of the line.) The average and half 
the difference of these results is used to determine the central value and uncertainty respectively.
}
\label{fig:ell0a4}
\end{center}
\end{figure}
We show the extrapolations for $F_3^\infty$ and $\cL(0)$ in 
Figs.~\ref{fig:f3infa4} and \ref{fig:ell0a4}, respectively.
Note that here the use of finite volume is simply a tool to discretize the equations,
and is not related to the volume of any simulation. We have worked up to $L=100$,
which, as the figures show, is enough to provide reasonable control over
the extrapolation.
For $F_3^\infty$, the results show oscillations for $L\lesssim 40$,
but for larger $L$ settle to a constant value.
We estimate the infinite volume value by fitting the large $L$ results to a constant.
For $\cL(0)$ we take the average of the linear and quadratic fits as the central value,
and half the difference as the error. We find
\begin{equation}
F_3^\infty = 4.0068(1) \times 10^{-5} \,,\qquad
\cL(\vec 0) = 0.276203(7)\,,   \qquad (a=0.41315)\,.
\end{equation}
Inserting these results into Eq.~(\ref{eq:MdftoKdfthr}) we obtain
\begin{equation}
\Kiso=10 \ \ \Longrightarrow\ \ \cM_{3,\df,\thr} =6 .8633(1)\,,  \qquad (a=0.41315)\,.
\end{equation}

We now turn to the relation between $\Mthr$ and $\cM_{3,\df,\thr}$.
The latter is given by
\begin{align}
\Mthr & =\Mdfthr + \tilde I_1 + \tilde I_2 + S_I\,,
\label{eq:MthrfromMdf}
\end{align}
where $\tilde I_1$, $\tilde I_2 $ and $S_I$ are defined in Appendix.~\ref{app:impintthr}. 
In all cases the quantities are obtained by taking infinite-volume limits of 
appropriate finite-volume quantities. As above, this is only a tool for discretizing integral equations and the parameter $L$ used here does not correspond to the finite-volume of the system.

\begin{figure}[tbh]
\begin{center}
%\vskip -.2truein
%\includegraphics[scale=0.8]{I1}
\includegraphics[width=0.75\textwidth]{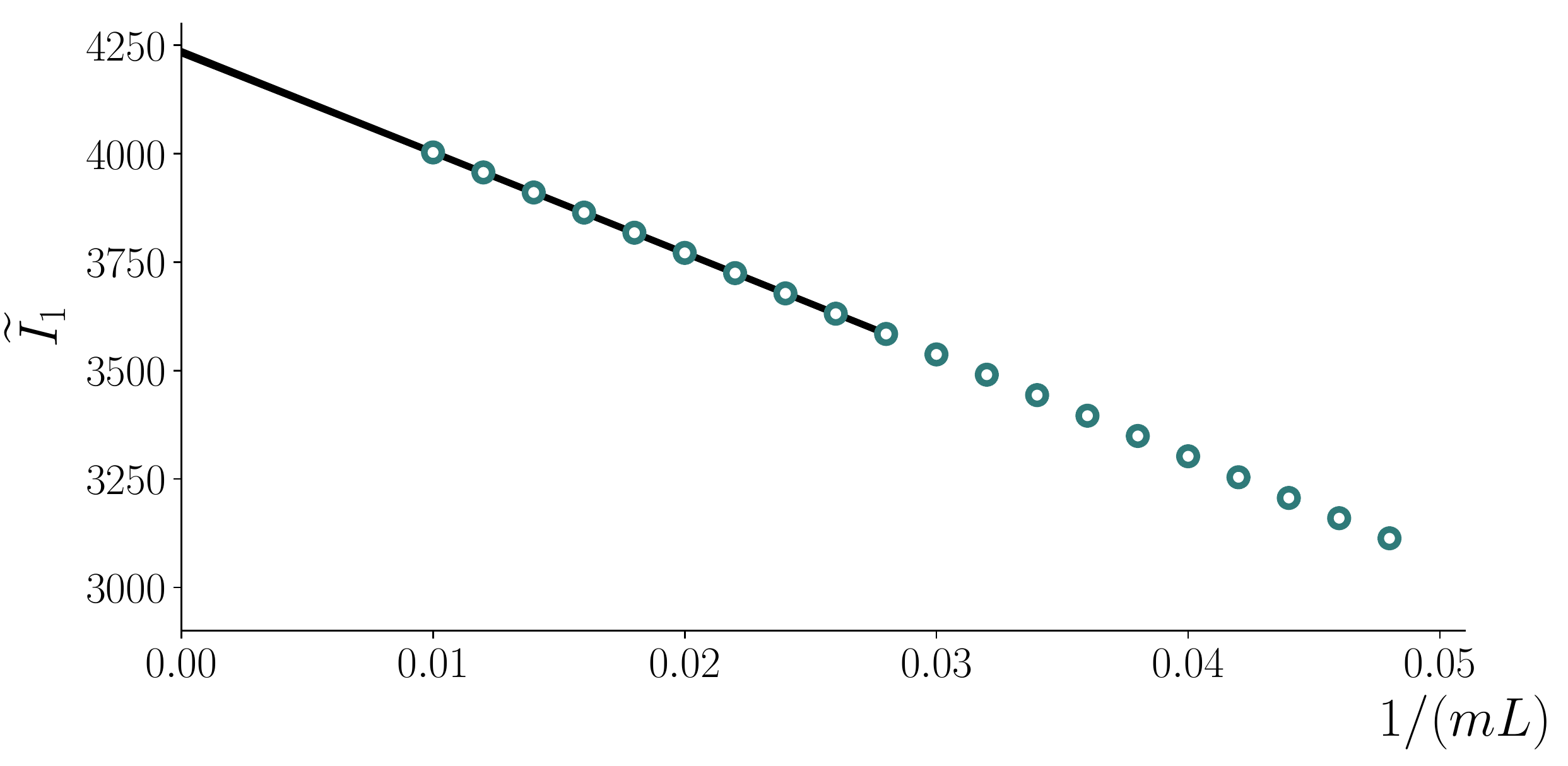}
%\vskip -4.2truein
\caption{
$\wt I_1$ for $ma=0.41315$ plotted versus $1/(mL)$, together with linear and quadratic fits to the range indicated by the curves. (Both curves are plotted but are indistinguishable aside from a thickening of the line.)  The average of the two $L\to \infty$ extrapolations is used as the central value and half the difference as the uncertainty.
}
\label{fig:I1}
\end{center}
\end{figure}
\begin{figure}[tbh]
\begin{center}
%\vskip -.2truein
%\includegraphics[scale=0.8]{I2}
\includegraphics[width=0.75\textwidth]{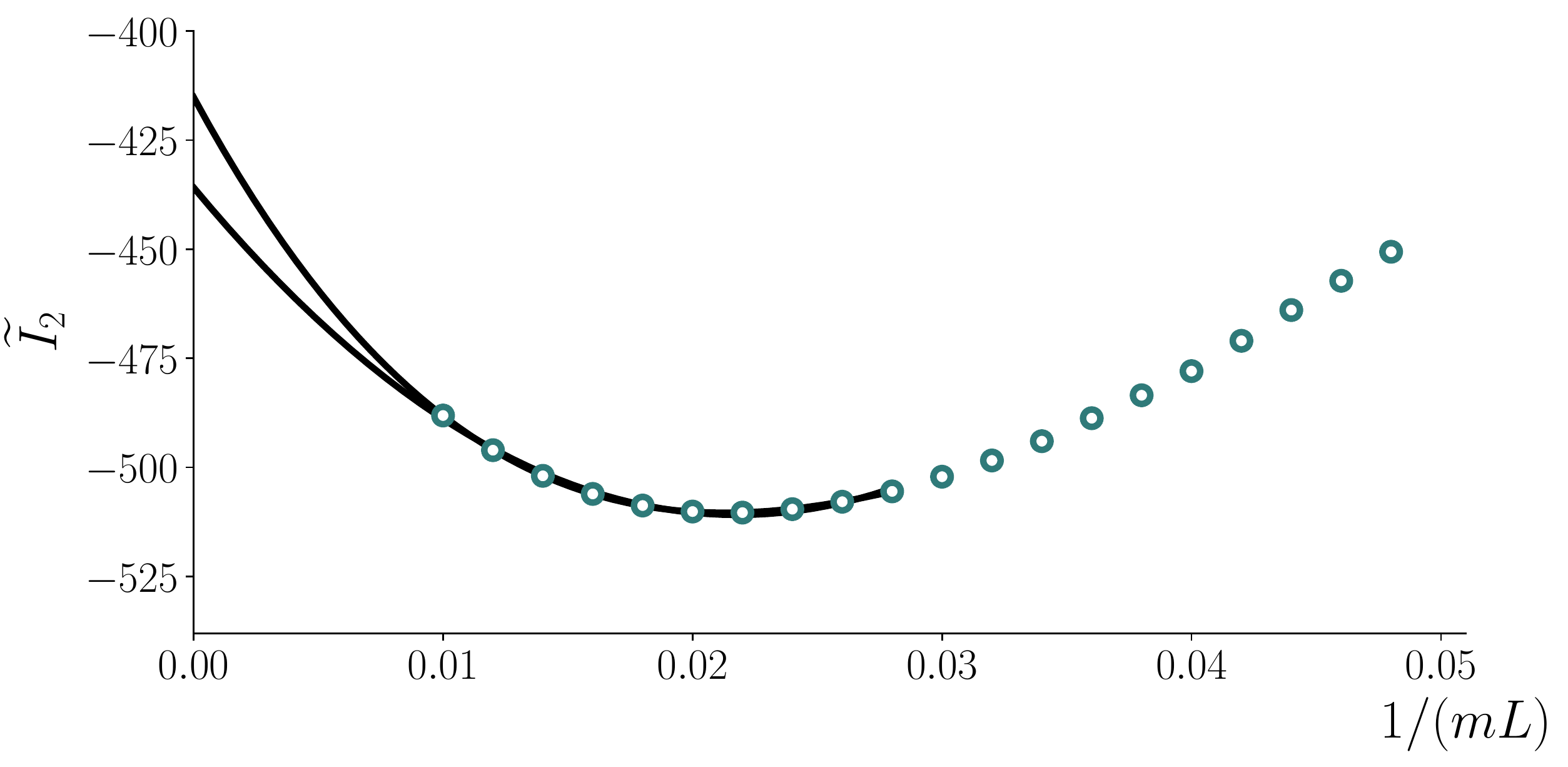}
%\vskip -4.2truein
\caption{
$\wt I_2$ for $ma=0.41315$ plotted versus $1/(mL)$,  together with quadratic and cubic fits to the range indicated by the curves.  The average of the two $L\to \infty$ extrapolations is used as the central value and half the difference as the uncertainty.
}
\label{fig:I2}
\end{center}
\end{figure}
\begin{figure}[tbh]
\begin{center}
%\vskip -.2truein
%\includegraphics[scale=0.8]{SI}
\includegraphics[width=0.75\textwidth]{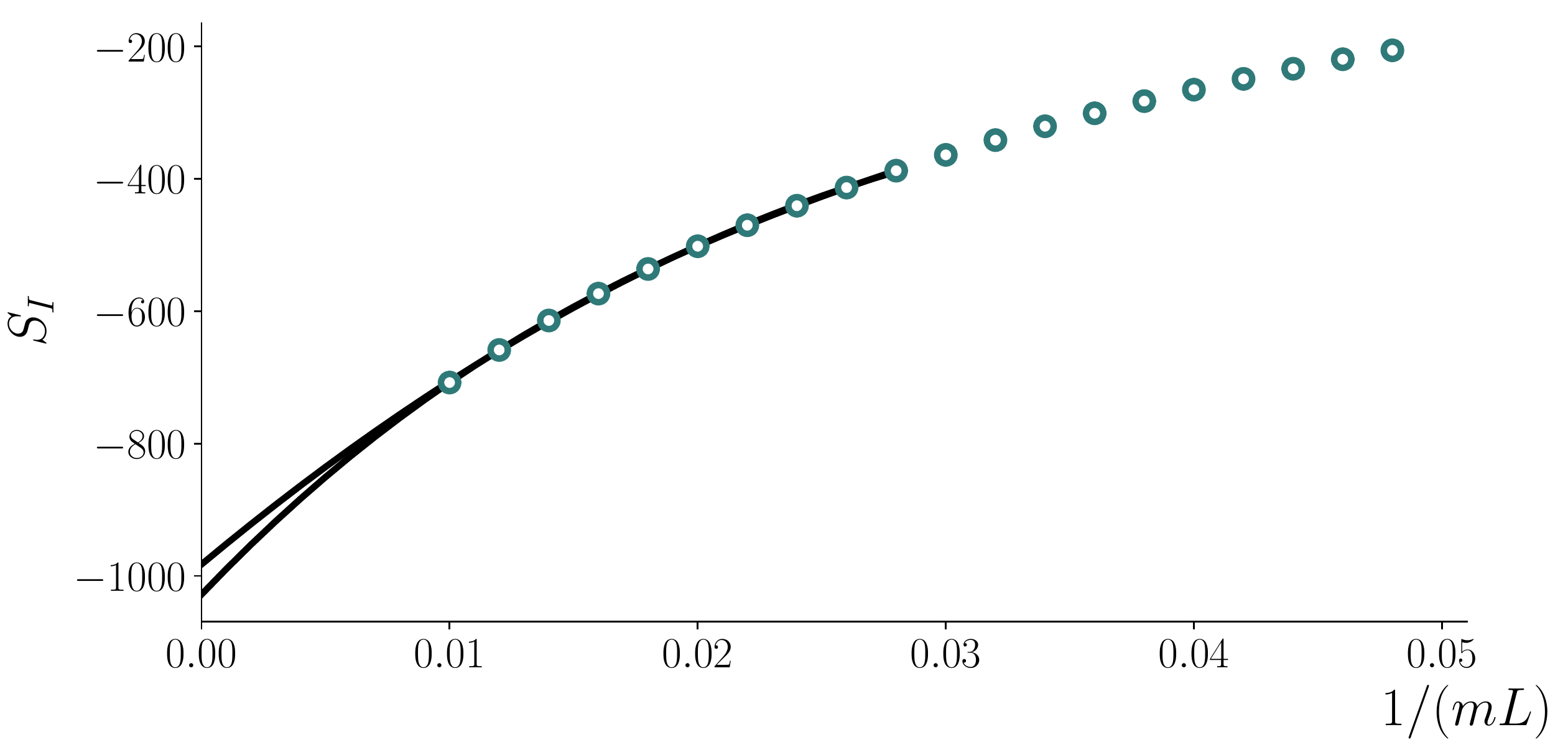}
%\vskip -4.2truein
\caption{
$S_I$ for $ma=0.41315$ plotted versus $1/(mL)$, together with quadratic and cubic fits to the range indicated by the curves. The average of the two $L\to \infty$ extrapolations is used as the central value and half the difference as the uncertainty.
}
\label{fig:SI}
\end{center}
\end{figure}

Results for $\wt I_1$, obtained using
Eq.~(\ref{eq:I1num}) are shown in Fig.~\ref{fig:I1}.
Values of $L$ up to 100 are easily attained, and the extrapolation to $L=\infty$ is 
well controlled. We find
\begin{equation}
\wt I_1 = 4233 \pm 2\,, \qquad (a=0.41315)\,.
\end{equation}
The corresponding extrapolation for $\wt I_2$, based
on Eq.~(\ref{eq:I2num}), is shown in Fig.~\ref{fig:I2}.
Our result for the infinite-volume limit is 
\begin{equation}
\wt I_2 = -425 \pm 10\,,  \qquad (a=0.41315)\,.
\end{equation}
Finally, the extrapolation leading to $S_I$ is shown in Fig.~\ref{fig:SI}, 
based on Eq.~(\ref{eq:SInum}), yielding
\begin{equation}
S_I =  -1005 \pm 23\,,  \qquad (a=0.41315)\,.
\end{equation}
Combining these results we find that
\begin{equation}
\Mthr-\cM_{3,\df,\thr} = 2803 \pm 25\,,  \qquad (a=0.41315)\,,
\end{equation}
where $\wt I_1$ dominates the overall value while $\wt I_2$ and $S_I$ dominate the error.
This shift dominates the value of $\cM_{3,\df,\thr} = 6.9$ found above.
The final result is thus
\begin{equation}
\frac{\Mthr}{48} = 58.5 \pm 0.5 \,.
\end{equation}
This is in good agreement with $\Mthr/48= 60.0 \pm 0.8$, 
the indirect value found using
the threshold expansion Sec.~\ref{sec:threxp}. This provides an important
cross-check on our calculations and formalism.

\bigskip

Finally, we calculate the dependence of $\Mthr$ on $\Kiso$, in order to compare to
the results obtained using the threshold expansion.
Noting that the relation between $\Mthr$ and $\cM_{3,\df,\thr}$ is independent
of $\Kiso$ we find, using Eq.~(\ref{eq:MdftoKdfthr}), that
\begin{equation}
-\frac{1}{48}\frac{\partial \Mthr}{\partial (1/\Kiso)} \bigg \vert_{a,E=3}
=  -\frac{1}{48}\frac{\partial \cM_{3,\df,\thr}}{\partial (1/\Kiso)} \bigg \vert_{a,E=3}      = \frac{1}{48}  \frac{9\cL( 0)^2}{(1/\Kiso + F_3^\infty)^2} \,.
\label{eq:dM3dfthr}
\end{equation}
Since we have determined $\cL( 0)$ and $F_3^\infty$, we can
immediately calculate this quantity. We plot the result
in Fig.~\ref{fig:derivmthrf} above as the solid line. The uncertainty in
this line from the volume extrapolation, which comes dominantly from
the uncertainty in $\cL( 0)$, is less than the width of the curve.
In the figure we compare this result
to that obtained above using the threshold expansion and find good agreement.

\section{Unphysical Solutions\label{sec:unphys}}

\begin{figure}
\begin{center}
\includegraphics[width=0.99\textwidth]{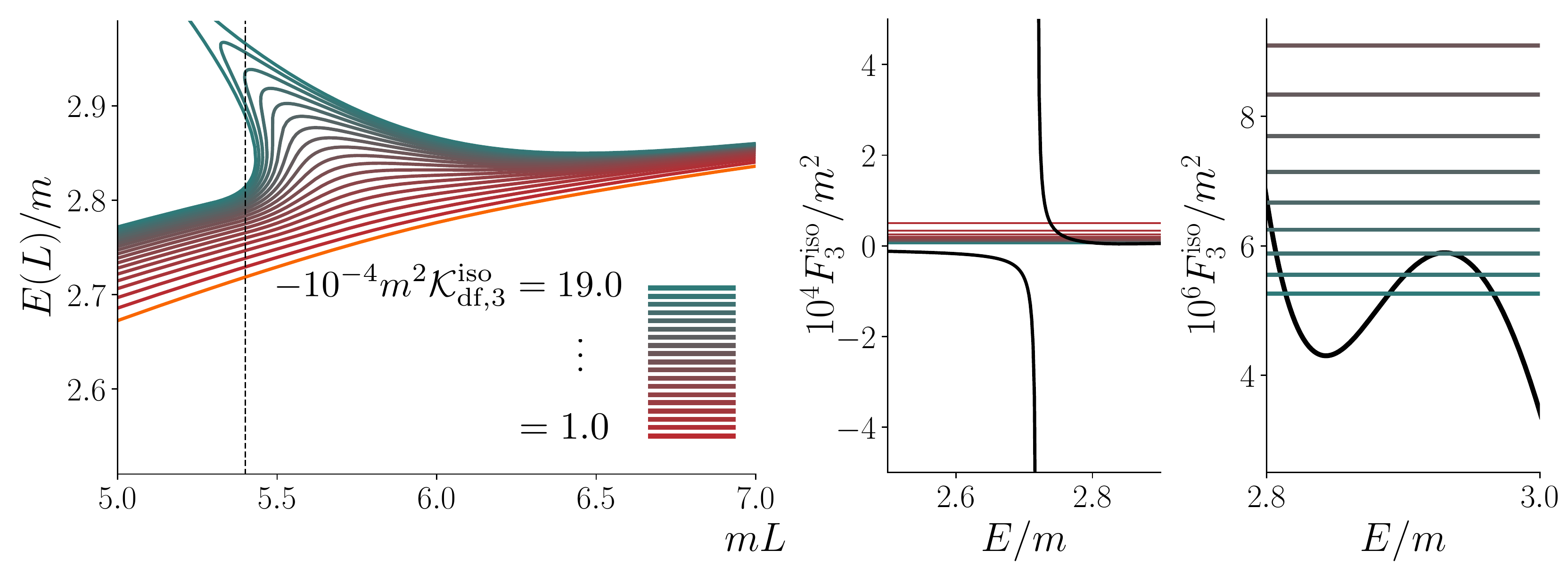}
\caption{
Plot showing an example of the unphysical solutions that arise for certain choices
of parameters, here for $ma=-10$ and large, negative values of $\Kiso$. 
The left panel shows the lowest finite-volume level,
with $-10^{-4} m^2 \Kiso$ ranging from 1 to 19 in unit steps.
This level also appears in Fig.~\ref{fig:negKdf}, but here we extend the results to more
negative values of $\Kiso$.
The unphysical behavior is the doubling back of the spectrum, so that there are three, rather
than one, levels for a range of $mL$ around 5.4 (the value shown by the vertical dashed line).
To understand how this doubling back arises, we show in the right two panels plots
of $F_3^{\text{iso}}/m^2$ vs.~$E/m$ for $mL=5.4$, 
together with the values of $-1/(m^2 \Kiso)$ whose intersections give the solutions. 
The middle panel looks reasonable, but the enlargement shown in the right panel 
reveals that $F_3^\iso$ is not decreasing monotonically, leading to the triplet of solutions
for the largest three values of $\vert\Kiso\vert$.
As explained in the text, the middle solution corresponds to a pole with
an unphysical residue. }
\label{fig:unphys}
\end{center}
\end{figure}

In this section we briefly discuss unphysical solutions that we have identified numerically 
when solving the
quantization condition for certain values of $a$ and $\Kiso$. 
As we explain below,
to guarantee physical solutions for all choices of $a$ and $\Kiso$,
$F_3^{\text{iso}}(E,L,a)$ must be a monotonically decreasing function of $E$. 
However, we find that this condition is violated for certain parameter choices. 
An example is shown in Fig.~\ref{fig:unphys}.

The monotonicity condition on $F_3^\iso$ is derived as follows.
We consider the finite-volume correlator
\begin{equation}
\label{eq:corrdef}
C_L(E) = \int_L d^4 x\; e^{ i E t} 
\langle 0 \vert \text{T} \mathcal O(x ) \mathcal O^\dagger (0) \vert 0 \rangle_L \,,
\end{equation}
where $\cO^\dagger(x)$ is any operator that creates three-particle states.
This is the correlator that is used in the derivation of the quantization 
condition in Ref.~\cite{\HSQCa}.
From Eq.~(42) of that work, we find that, when restricted to the isotropic approximation,
the correlator in the vicinity of a pole has the form
\begin{equation}
C_L(E)  = i A(E) \frac{1}{\Kiso(E) + F_3^{\text{iso}}(E,L,a)^{-1}} A^*(E) + \textrm{regular}\,,
\label{eq:CLE}
\end{equation}
where $A(E)$ is an infinite-volume matrix element connecting the vacuum to a three-particle
state by the action of $\cO$.
From the spectral decomposition of $C_L(E)$, we know that the pole has the form 
\begin{equation}
C_L(E) = i \frac{c}{E-E_n(L)} + \textrm{regular}
\,,
\end{equation}
with $E_n(L)$ the pole position and $c$ a positive, real constant.
For this to be true for the pole in Eq.~(\ref{eq:CLE}), the following condition must
be satisfied 
\begin{equation}
\label{eq:physcond}
\left[\frac{\partial F_3^{\text{iso}}(E,L,a)}{\partial E} 
+ \frac{\partial 1/\Kiso(E)}{\partial E} \right]_{E=E_n(L)}  < 0\,.
\end{equation}
For a constant $\Kiso$, as used in most of this work and,
specifically, in Fig.~\ref{fig:unphys}, this implies that $F_3^\iso$ must be decreasing
at the crossing point.
Assuming that there are physical theories with all values of $\Kiso$,
the crossing point can occur anywhere along the curve, and thus
$F_3^\iso$ must be monotonically decreasing.
Assuming that all values of $a$ are physical, this monotonicity property must hold
in general.

We now return to Fig.~\ref{fig:unphys}.
This shows an example where $F_3^\iso$ does not decrease monotonically with $E$,
but instead, as shown in the right panel, has a small upward excursion.
This implies that, in a small range of $\Kiso$ there are three solutions to the
quantization condition, the middle of which violates the condition 
Eq.~(\ref{eq:physcond}). To obtain three states, the spectral curves
must double back, as shown in the left panel.
Thus this doubling back is an alternative criterion for unphysicality.

Clearly the appearance of such solutions is problematic and needs to
be understood. This is work in progress, but based on our tests so far
we can offer some remarks. We find that $F_3^{\text{iso}}$ only
develops a positive slope in regions where its magnitude is small and,
as the volume is increased, these regions always go away and the
function becomes ``healthy''. This leads us to suspect that we are seeing
a form of neglected finite-volume effects that are formally
exponentially suppressed but with oscillatory energy dependence. These
cause problems when small values of $F_3^{\text{iso}}$ are sampled by 
large-magnitude values of $\Kiso$.

In addition, the oscillations in $F_3^{\text{iso}}$ share similarities with
the oscillations observed in the threshold state of
Fig.~\ref{fig:thra04}, and seem to be connected. 
In that case we found that using a different
definition of $\widetilde F^s$ removed the oscillations and this
points to the fact that the smooth cutoff function, $H(\vec k)$, may be
the source of the issue. This is plausible because, although smooth,
the cutoff function does have vanishing support above a certain value
of $k$. It is well known that sharp cutoffs lead to oscillatory
behavior, and the oscillations here might be a related phenomenon.

It is also possible that the unphysical solutions are an indication that
the isotropic, low-energy approximation is breaking down for certain
choices of parameters.
Our approach for now is to avoid the (relatively small) regions in which 
unphysical solutions occur, while at the same time actively investigating their source. 

\bigskip

We close with some more general remarks about the condition Eq.~(\ref{eq:physcond}).
First, we stress that, since $\Kiso$ can depend on $E$, one should in general
use the full condition including the derivative of $1/\Kiso(E)$.
Second, we note that a similar condition holds for the two-particle quantization condition,
in the s-wave approximation, namely
\begin{equation}
\label{eq:physcond2}
\left[
\frac{\partial F_s(E,\vec k,L)}{\partial E} 
+ \frac{\partial 1/\cK_2^s(\vec k)}{\partial E} \right]_{E=E_n(L)}  < 0\,,
\end{equation}
where $F_s = 2 \omega_k \wt F_s(\vec k)$, $E$ is the total two-particle energy,
and $E_n(L)$ is here a solution to the two-particle quantization condition,
$F_s = - 1/\cK_2^s$. Note that here we are considering also a moving frame,
with $- \vec k$ being the total momentum.
The result (\ref{eq:physcond2}) can be derived using 
the result of Ref.~\cite{Kim:2005gf} for the two-particle correlation function,
following similar steps to those outlined above.
To our knowledge it has not been presented before.
The solutions to the two-particle quantization condition
shown in the left panel of Fig.~\ref{fig:QCexample} all satisfy this condition.

We can use Eq.~(\ref{eq:physcond2}) to learn about the way in which such
consistency conditions can fail. 
In all examples that we are aware of, $F_s$ is  a monotonically decreasing
function of $E$.
Thus a violation of this condition requires $1/\cK_2^s$ to rise sufficiently rapidly
at the crossing point. If this is the case, there will be spectral lines that double back,
similar to those shown in Fig.~\ref{fig:unphys}.
If this occurs for some $L$, the problem will go away as the box size increases, 
because $F_s$ becomes an increasingly steep function of $E$ as $L$ is taken larger.
In this case it seems that there are two possible causes for the problem.
One is that it is caused by neglected exponentially suppressed corrections,
the other that the choice of rapidly increasing $1/\cK_2^s$ is simply unphysical
within the s-wave approximation.
The problem cannot, however, lie with the $H$ functions, since $F^s$ can be
regulated in other ways.

Our third observation is that the consistency condition,
Eq.~(\ref{eq:physcond2}), 
turns into the condition introduced in Ref.~\cite{Iritani:2017rlk} 
if a subthreshold solution persists as $L \to \infty$.
This is because $F_s \to \rho$ in this limit, and one
can then show algebraically that the conditions are equivalent. 
This is as expected since the persistence of a subthreshold solution is equivalent to the 
existence of a bound state,
and Eq.~(\ref{eq:physcond2}) is just the requirement that this pole,
which remains isolated as $L \to \infty$, has a residue with the proper sign.  
 
Finally, we can relate Eq.~(\ref{eq:physcond2}) to a result from Ref.~\cite{Hansen:2012tf}.
In that work it was noted that Lellouch-L\"uscher factors were physical  only 
if the condition $\partial(\delta_s+\phi^P)/\partial E^* > 0$ was satisfied, where $\delta_s$ is the
s-wave phase shift and $\phi^P$ is the L\"uscher pseudo-phase, related to $\widetilde F^s$.
It is straightforward to show that this condition is equivalent to Eq.~(\ref{eq:physcond2}). 
We note also that, from the perspective of Refs.~\cite{Briceno:2015csa,Briceno:2014uqa},
this equivalence is clear since the Lellouch-L\"uscher-like relation is derived there
via the same matching condition that leads to Eq.~(\ref{eq:physcond2}) here.

\section{Conclusions and Outlook\label{sec:conclusion}}

In this work we have numerically explored the relativistic three-particle quantization condition derived in Refs.~\cite{\HSQCa,\HSQCb}. In order to capture the key features of the formalism, and to compare the work flow to that described in Ref.~\cite{\Akakib},
we have made a number of simplifications. 
Specifically we have restricted attention to vanishing total momentum in the box, 
truncated the infinite-volume scattering observables to the s-wave, isotropic approximation,
and taken the two-particle sector to be dominated by the scattering length in the effective-range expansion.

Within this reduced set-up, we find that the quantization condition is numerically straightforward to implement and that a great deal of interesting physics is buried in the simple formula. For example, as summarized in Fig.~\ref{fig:EvsLVaryA}, the condition provides a useful benchmark, by predicting the part of the volume-dependence of three-particle energies that is due only to two-particle scattering---i.e.~the case where the three-particle contact interaction is neglected, $\Kiso=0$. This is a useful starting point in lattice calculations since infinite-volume, three-particle scattering information can only be recovered by measuring deviations from these benchmark curves.

Going beyond this, we show how turning on nonzero values of $\Kiso$ predicts a rich set of phenomena, including resonance-like avoided level crossings [Fig.~\ref{fig:res}] and the finite-volume energy shift for an Efimov-like three-particle bound state [Fig.~\ref{fig:BS1}]. The latter case is particularly clean as we can study the state over a vast range of volumes, $mL = 4$ to $70$, and show that the predicted level matches the asymptotic predictions for $\kappa L \gg 1$ (in our case implying $m L \gg 10$), but also show how the level deviates from the asymptotic form and thus that the full formalism is needed to make reliable predictions for realistic volumes. Finally, our result also describes the regime of weak interactions and, as we discuss in Sec.~\ref{sec:threxp}, we can numerically resolve all known terms in the $1/L$ expansion of the threshold state, including the $\log(mL)/L^6$ dependence.

Beyond predicting detailed finite-volume behavior, our formalism also provides a powerful tool in understanding the infinite-volume scattering of three-particle states. This is because a simple form of $\Kiso$ corresponds to a complicated three-particle scattering amplitude, $\mathcal M_3$, with nontrivial phase space dependence generated dynamically by the integral equations relating $\Kiso$ to $\mathcal M_3$. The most dramatic example of this is summarized in Fig.~\ref{fig:BoundGammavsk}, where we take two inputs designed to produce a shallow bound state ($\Kiso=2500$ and $a=-10^4$) and from this predict the wave-function with no free parameters. The numerical reproduction of this complicated functional form, which spans many orders of magnitude, gives us confidence that the approach of relating $\Kdf$ to $\mathcal M_3$ should be a useful tool in describing three-particle physics for a variety of systems.

Despite these successes, future work is needed to bring this formalism to maturity for use in realistic numerical LQCD calculations. In this direction it is instructive to first compare our approach to that using NREFT, described in Refs.~\cite{\Akakia,\Akakib}. One key difference between our formalism and the NREFT proposal is that the latter uses a hard cutoff in place of our smooth cutoff function $H(\vec k)$, and places this cutoff at much higher spectator momenta. 

Recalling that $(E, \vec P)$ is fixed, the spectator momentum  $(\omega_k, \vec k)$ determines the invariant mass squared of the nonspectator pair to be $E_{2,k}^{*2} = (E - \omega_k)^2 - (\vec P - \vec k)^2$. Thus, taking $\vec k$ very large takes $E_{2,k}^{*2}$ not only below $4m^2$ but in fact to negative values with large magnitude. From the perspective of our approach, dependence on the deep sub-threshold values of the two-to-two scattering amplitude is undesirable, leading us to introduce $H(\vec k)$. 
{Among other things, this avoids the region of the left-hand cut in the
two-to-two amplitude. 
This region is accessed in the approach of Refs.~\cite{\Akakia,\Akakib}, 
but the left-hand cut is avoided by restricting the NR expansion to a few terms.

We emphasize the role of $H(\vec k)$ once more here, because we suspect this to be related to unphysical finite-volume energies that we find for certain values of $\Kiso$. As we describe in Sec.~\ref{sec:unphys}, the finite-volume function $F_3^{\text{iso}}$ is generally monotonically decreasing with energy, but can exhibit small upward oscillations for volumes up to $mL \approx 6$, i.e.~including nearly all present-day lattice calculations. These oscillations lead to unphysical solutions when $|\Kiso|$ is large enough to intersect them. 

Understanding the exact nature of these artifacts and modifying the formalism to remove them is clearly crucial. As a first step we note that varying the width of the cutoff function, $H(\vec k)$, can show which regions suffer from these effects. Thus one can identify values of $F_3^{\text{iso}}$ where the artifacts do not arise and restrict attention to values of $\Kiso$ that only intersect these regions. This is only a first step as our ultimate goal is a formalism that works for all possible scattering parameters, with no need to identify safe regions numerically.

In addition to addressing the issues mentioned above, future projects include going beyond the isotropic approximation in a systematic way, including the role of the mixing of different angular-momentum states in finite volume. Such mixing is already built into our full quantization condition so the task is one of block-diagonalization, or subduction, onto the irreps of the finite-volume symmetry group. Additional formal steps include incorporating $\mathcal K_2$ poles, multiple two- and three-particle channels and nonidentical and nondegenerate particles into our formalism, as well as particles with intrinsic spin. Finally, on the side of infinite-volume physics, we are in the process of developing tools to numerically relate $\Kdf$ and $\mathcal M_3$ also above threshold. Here it will be crucial to develop realistic parametrizations of three-particle scattering amplitudes, especially in resonant channels.

\section*{Acknowledgments}

The work of SRS was supported in part by the United States Department of Energy grant No.~DE-SC0011637. RAB acknowledges support from U.S. Department of Energy contract DE-AC05-06OR23177, under which Jefferson Science Associates, LLC, manages and operates Jefferson Lab. 
We thank Akaki Rusetsky, as well as all other participants of the INT workshop ``Multi-Hadron Systems from Lattice QCD'' (February 5 - 9, 2018), for useful discussions,
and Tyler Blanton for comments on the manuscript.

\appendix

\section{Numerical implementation}
\label{app:numerics}

The numerical implementation naturally falls into two parts: (i) applying the 
finite-volume quantization condition Eq.~(\ref{eq:QCiso}), 
and (ii) solving the infinite-volume integral equation Eq.~(\ref{eq:Duu}) and doing the integrals
in Eqs.~(\ref{eq:cL})-(\ref{eq:wtrho}). 
We consider these parts in turn.
As in the rest of this work, it is convenient to use units in which $m=1$.
Factors of $m$ can be added back using dimensional analysis.

\subsection{Implementing the quantization condition}
\label{app:numQC}

{The matrices $\widetilde F^s$, $\widetilde G^s$, etc.,~entering the quantization condition
(\ref{eq:QCiso}) are all of size $N\times N$.
Here $N$ is determined by the cutoff function $H(\vec k)$.
For given choices value of $E$, $L$ and $\alpha$, 
$H(\vec k)$ is nonzero only for a finite number of finite-volume momenta $\vec k$. }
For example, if $E=4$ then $N=19$, $93$ and $895$ for
$L=5$, $10$ and $20$, respectively.

To simplify the numerical computation, and, in particular, to allow a straightforward
determination of the dependence on $a$, we rewrite $F_3^s$ as
\begin{align}
\left[F_3^s\right]_{kp} &= \frac1{L^3} \left[
\frac{\widetilde  F^s}3 - \frac{\widetilde F^s}{\zeta} 
\left(\frac1{H_{FG}-1/a}\right)
\frac{\widetilde F^s}{\zeta}\right]_{kp}\,,
\label{eq:F3sa}
\\ 
 \zeta_{kp} & = \delta_{kp} \frac{1}{\sqrt{32 \pi \omega_k E_{2,k}^*}}\,,
\label{eq:zeta}
\\
\left[H_{FG}\right]_{kp} &= \delta_{kp} |q_{2,k}^*|[1-H(\vec k)]
+
\left[\frac{\widetilde F^s}{\zeta^2} + \frac1{\zeta} \widetilde G^s \frac1{\zeta}
\right]_{kp}
\label{eq:HFG}
\,.
\end{align}
$H_{FG}$ is a real, symmetric matrix that can be diagonalized as
\begin{equation}
H_{FG} = \sum_{n=1}^N |n\rangle \lambda_n \langle n |
\,,
\label{eq:HFGdiag}
\end{equation}
where $\lambda_n$ and $ |n\rangle$ are its eigenvalues and eigenvectors respectively. To implement the sums over $\vec k$ and $\vec p$ in the definition of $F_3^\iso$,
Eq.~(\ref{eq:F3iso}), we use the vector $|\mathbf 1\rangle$ introduced following 
Eq.~(\ref{eq:QCiso}) in the main text.
Putting this together we find
\begin{equation}
F_3^\iso =
\frac1{L^3}\left\{
\frac{\langle \mathbf 1| \widetilde F^s | \mathbf 1\rangle }3
- 
\sum_n \frac{\langle\mathbf 1 | {\widetilde F^s}/{\zeta} \,| n\rangle^2 }
{-1/a+\lambda_n}
\right\}
\,.
\label{eq:finalF3iso}
\end{equation}
Thus, in order to determine $F_3^\iso$, it is convenient to construct $H_{FG}$,
then diagonalize it, and finally calculate the (real) matrix elements
$\langle \mathbf 1| \widetilde F^s | \mathbf 1\rangle$
and
$\langle\mathbf 1 | {\widetilde F^s}/{\zeta} \,| n\rangle$.
Given the eigenvalues and these matrix elements we know $F_3^\iso$ 
(at the chosen values of $E$ and $L$) for all values of $a$.
An example of the $a$ dependence is shown in Fig.~\ref{fig:F3isoE4L10} where the energy and volume have been fixed to $E=4$ and $L=10$ respectively.
Because of the overall factor of $1/L^3$, $F_3^\iso$ has a small magnitude
except near the poles at $a=1/\lambda_n$. The figure shows a typical example
in that most of the poles are in the region $a\ge 1$ where our formalism does not hold.

\begin{figure}[tbh]
\begin{center}
%\vskip -.2truein
\includegraphics[width=0.75\textwidth]{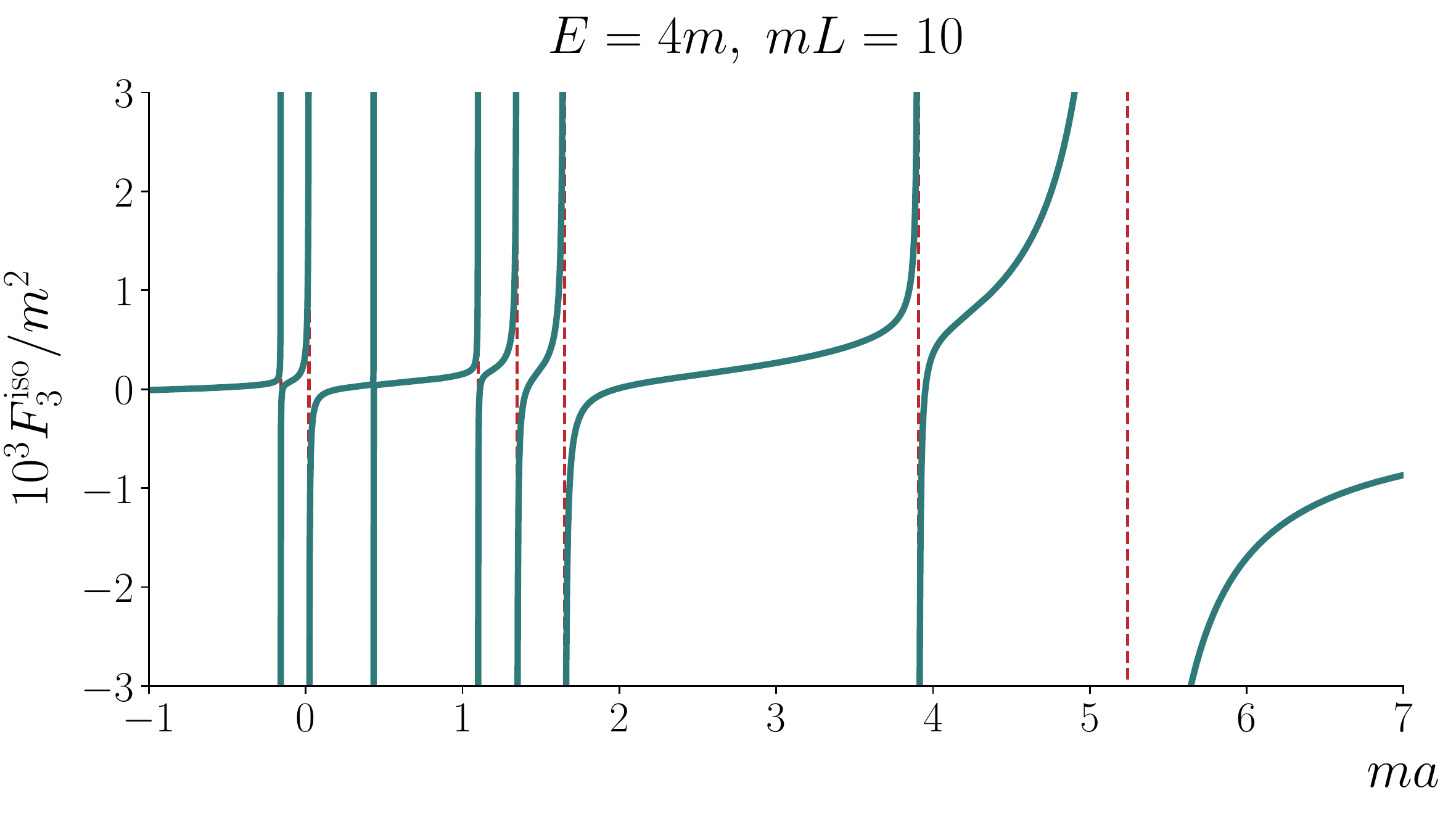}
%\vskip -4.2truein
\caption{
$F_3^\iso/m^2$ vs.~$ma$ for $E=4m$ and $mL=10$. 
There are eight poles in total (shown by vertical [red] dashed lines), 
one per momentum shell.
}
\label{fig:F3isoE4L10}
\end{center}
\end{figure}

We can substantially reduce the size of the matrices needed in the calculation using
group theory~\cite{\HSQCa}.
The finite-volume momenta fall into ``shells'',
the members of which are related by elements of the octahedral group $O_h$ (which we
define to include parity). For example, the first four shells have 1, 6, 12 and
8 members, respectively, with representative elements being $2\pi \vec n/L$
with $\vec n=(0,0,0)$, $(0,0,1)$, $(0,1,1)$ and $(1,1,1)$.
The matrices $\wt \rho$, $\wt F^s$ and $\wt G^s$ are invariant under
cubic group transformations, implying that the eigenvectors of $H_{FG}$ lie
in irreducible representations (irreps) of the group.
The state $|\mathbf 1\rangle$ projects each shell onto the fully symmetric 
$A_1^+$ irrep, and the invariance of $\wt F^s$ and $\xi$ implies that
only eigenvectors $|n\rangle$ lying in the $A_1^+$ irrep contribute to $F_3^\iso$.
The net result is that we need only invert the $A_1^+$ block of $H_{FG}$,
which is an $N_{\rm sh}\times N_{\rm sh}$ matrix, with $N_{\rm sh}$ being the
number of shells for which $H(\vec k)\ne 0$.
This drastically reduces the matrix size and concomitantly speeds up the numerical
evaluation. For the examples given above, where $E=4$, and $L=5$, $10$ and $20$,
the values of $N_{\rm sh}$ are $3$, $8$ and $40$
(compared to $N=19$, $93$ and $895$). 
Note also that the $8$ poles in Fig.~\ref{fig:F3isoE4L10} directly correspond 
to the $8$ shells for $E=4$, $L=10$. Although we focus on systems at rest here, 
one may speed up calculations for systems with nonzero total momenta by consider 
the irreps of the little groups of $O_h$.

$\wt F^s$ is our version of the zeta-function introduced by L\"uscher in
Refs.~\cite{\Luscher}. We give a brief description of the methods we use to calculate
it in Appendix \ref{app:Fs}, below.

\subsection{Implementing the integral equations and integrals below threshold}
\label{app:impint}

As discussed in Sec.~\ref{sec:KtoM}, in order to relate $\Kiso$ to physical quantities, it is general necessary to solve an integral equation, Eq.~(\ref{eq:Duu}). In this study we have restricted our attention to energies below and at threshold, where this procedure is relatively straightforward. Here we explain how the infinite-volume limit can be achieved for these kinematics.
The below-threshold case is simplest and
we discuss it first. In particular, we are interested in determining the quantities
appearing in the equation for $\Mdf$, Eq.~(\ref{eq:M3dKiso}).
In particular, there will be a bound state whenever
$F_3^\infty=-1/\Kiso$, with $\cL(\vec k)=\cR(\vec p)$ 
being proportional to the on-shell Bethe-Salpeter wavefunction of this bound state.

The determination of $F_3^\infty$ below threshold is straightforward: 
we can simply take the $L\to\infty$ limit of our numerical evaluations of $F_3^\iso$,
i.e.
\begin{align}
F_3^{\infty} & =  \lim_{L \to \infty} F_3^{\text{iso}}
 =  \lim_{L \to \infty}  \sum_{\vec k, \vec p} \left[F_3^s\right]_{kp} \,.
\label{eq:F3inflim} 
\end{align}
This is an example of the limit employed in Ref.~\cite{\HSQCb} in order to relate the 
finite- and infinite-volume scattering amplitudes. 

To explain why Eq.~(\ref{eq:F3inflim}) is valid, we first note that, as explained in Ref.~\cite{\HSQCb}
\begin{equation}
\wt F^s(\vec k) \xrightarrow{L\to\infty} \wt \rho(\vec k)
\,.
\label{eq:Ftorho}
\end{equation}
This holds for all values of $E$, but we are interested only in $E < 3$, where
$\wt\rho$ is real. Naively one would have expected the right-hand side
to vanish, but the nonzero result arises because we use the PV prescription
in the integral in $\wt F^s$.\footnote{%
This result also serves as a useful check of our numerical evaluation of $\wt F^s$.}
The expression for $F_3^s$, Eq.~(\ref{eq:F3s}), can thus be rewritten in the infinite-volume
limit as
\begin{align}
\left[F_3^s\right]_{kp} &\xrightarrow{L\to\infty}  
\frac{1}{L^3}\left[ \frac{\wt \rho}3 - \wt \rho
\frac1{1/(2\omega \cM_2^s) + \wt G^s} \ \wt \rho\right]_{kp} \,,
\label{eq:F3sinf}
\\
&=
\frac{1}{L^3}\left[ \frac{\wt \rho}3 - \wt \rho (2 \omega \cM_2^s) \wt \rho
+\wt \rho 
\frac1{1 + (2\omega \cM_2^s)\wt G^s} (2\omega \cM_2^s) 
\wt G^s (2\omega \cM_2^s) \wt \rho\right]_{kp} \,.
\label{eq:F3sinfa}
\end{align}
Here the new matrices are
\begin{align}
\wt \rho_{kp} = \delta_{kp} \wt \rho(\vec k) 
\ \ {\rm and} \ \
\left[2\omega \cM_2^s\right]_{kp} =
\delta_{kp} 2\omega_k \cM_2^s(\vec k) 
\,,
\end{align}
and to obtain Eq.~(\ref{eq:F3sinf}) we have used the definition of $\mathcal M_2^s$,
Eq.~(\ref{eq:M2s}).

Next we note that the integral equation for $\cD^{(u,u)}$,
Eq.~(\ref{eq:Duu}), can be discretized as
\begin{equation}
\cD^{(u,u)}_{kp} = -\left[L^3(2\omega \cM_2^s) \wt G^s (2 \omega \cM_2^s)
+ (2\omega \cM_2^s) \wt G^s \cD^{(u,u)} \right]_{kp}
\,.
\label{eq:Duudisc}
\end{equation}
Here $\cD^{(u,u)}_{kp}=\cD^{(u,u)}(\vec k, \vec p)$ for finite-volume momenta,
and we have used the definitions of $\wt G^s$ and $G^\infty$ in
Eqs.~(\ref{eq:Gs}) and (\ref{eq:Ginf}).\footnote{%
In general, to take the $L\to\infty$ limit of $\wt G^s$, we have to introduce
a pole prescription~\cite{\HSQCa}, but this is not the case below threshold, 
because there are no poles. Note that $\wt G^s$ does not come with a built-in
pole prescription, unlike $\wt F^s$.}
Since Eq.~(\ref{eq:Duudisc}) is now a finite matrix equation, its solution is
\begin{equation}
\cD^{(u,u)} = -L^3\frac1{1+ (2\omega \cM_2^s) \wt G^s}
(2\omega \cM_2^s) \wt G^s (2 \omega \cM_2^s)
\,.
\label{eq:Duudiscsol}
\end{equation}
Using this we can rewrite Eq.~(\ref{eq:F3sinfa}) as
\begin{equation}
F_3^s \xrightarrow{L\to\infty}
\frac{\wt \rho}{3 L^3}
- \frac1{L^3} \wt \rho (2\omega \cM_2^s) \wt \rho
- \frac1{L^6} \wt \rho\, \cD^{(u,u)} \wt \rho\,.
\label{eq:F3sinfb}
\end{equation}
Comparing to Eqs.~(\ref{eq:cL}) and (\ref{eq:F3inf}), and
using the fact that $1/L^3 \sum_k \to \int_{\vec k}$ as $L\to \infty$,
we find the claimed result, Eq.~(\ref{eq:F3inflim}).

By similar arguments, one can evaluate the Bethe-Salpeter amplitudes using
either of the forms
\begin{align}
\cL(k)=\cR(k)  &= \lim_{L \to \infty} 
\frac13 -
\sum_p \left[\frac1{1/(2\omega \mathcal M_2^s)+ \widetilde G^s}  \wt \rho\right]_{kp}
\,,
\label{eq:calLinf}
\\
& = \lim_{L \to \infty}    \sum_{ \vec p}  L^3 \left[  [\widetilde F^s]^{-1} F_3^s\right]_{kp} \,. \label{eq:calLinflim}
\end{align}

We stress that, when using the results Eq.~(\ref{eq:F3inflim})-(\ref{eq:calLinflim}), 
$L$ is no longer playing the role of the spatial volume in the lattice calculation.
Instead, it allows for a convenient discretization of integral equations.
In particular, we are here interested only in the limiting values as $L\to\infty$,
and not to the form of the finite-$L$ corrections.

\subsection{Implementing the integral equations and integrals at threshold}
\label{app:impintthr}

We now explain how we solve the integral equations and perform the
integrals when working directly at threshold. 
The only change compared to the subthreshold case is that
$\wt G^s$ has a pole when $\vec k=\vec p=0$, which leads to an
IR divergence in $\cD^{(u,u)}$.
However, this IR divergence is absent for all the quantities of interest, 
either because it is multiplied by $\wt\rho(\vec 0)$, which vanishes at threshold, 
or because it appears in an IR finite integral. 
Thus we can regularize in the IR simply by removing the single divergent entry in $\wt G^s$:
\begin{equation}
\wt G^s_{kp} \longrightarrow \slashed{G}_{kp} = 
\begin{cases}  0 & \vec k=\vec p=0\\
\wt G^s_{kp} & \textrm{otherwise}
\end{cases}
\,.
\end{equation}
The finite-volume version of $\cD^{(u,u)}$ is then given by
\begin{equation}
\slashed{\cD}^{(u,u)} = -L^3\frac1{1+ (2\omega \cM_2^s) \slashed{G}}
(2\omega \cM_2^s) \slashed{G} (2 \omega \cM_2^s)
\,,
\label{eq:Duuslash}
\end{equation}
which is simply Eq.~(\ref{eq:Duudiscsol}) with $\wt G^s$ replaced by $\slashed{G}$.
Now, since $\wt F^s \to \wt \rho$ even at threshold, we can still use
Eqs.~(\ref{eq:F3sinf}), (\ref{eq:F3sinfa}) and (\ref{eq:F3sinfb}), as long as 
$\wt G^s \to \slashed{G}$ and the quantity being calculated is IR finite.
This leads to the results (all at threshold)
\begin{align}
F_3^\infty &= 
\frac1{3L^3} \tr(\wt \rho) 
- \frac1{L^3}\tr(\wt \rho\, (2 \omega \cM_2^s)\, \wt \rho)
- \frac1{L^6} \sum_{kp} \left[\wt \rho \,\slashed{\cD}^{(u,u)} \wt \rho \right]_{kp} + \cO(1/L)\,,
\label{eq:F3infnum}
\\
\cL(\vec 0) &= \cR(\vec 0) =
\frac13 - \frac1{L^3} \sum_k \left[\slashed{\cD}^{(u,u)} \wt \rho\right]_{0k} + \cO(1/L)
\,.
\label{eq:L0num}
\end{align}
The quantities on the right-hand sides of these equations 
can be evaluated numerically by a slight extension of the
work needed to solve the quantization condition, and taking $L\to\infty$ gives the 
left-hand sides. These are then combined to determine  $\Mdfthr$
using Eq.~(\ref{eq:MdftoKdfthr}) in the main text.

Similar methods allow the numerical determination of the relation between
$\Mthr$ and $\cM_{3,\df,\thr}$ that is given in Eq.~(123) of Ref.~\cite{\HSTH}.
The basic relation is given in Eq.~(\ref{eq:MthrfromMdf}),
and we give here the definitions of the quantities $\wt I_1$, $\wt I_2$ and $S_I$ that
appear in that equation.

First, making uses of Eqs. (119), (123) and (126) of Ref.~\cite{\HSTH},
we find
\begin{equation}
\wt I_1 = 9 L^3
\left[ \left(2 \omega \cM_2 \slashed{G} \right)^2 2\omega \cM_2\right]_{00}
+ 9\times 2^{12} m^2 \pi^3 a^3
\frac{1}{L^3}\sum_{\vec k\ne0}
\left[\frac{H(\vec k)^2}{k^4} +a \frac{\sqrt3}2 \frac{H(\vec k)^3}{k^3}\right]
+ \cO(1/L) \,.
\label{eq:I1num}
\end{equation}
Both terms have linear and logarithmic divergences as $L\to\infty$, but these cancel
to leave a finite limit.
The corresponding result for $\wt I_2$,
obtained using Eqs.~(122), (123) and (125) of Ref.~\cite{\HSTH}, is
\begin{align}
\wt I_2 &= -9 L^3
\left[ \left(2 \omega \cM_2^s \slashed{G} \right)^3 2 \omega \cM_2^s\right]_{00}
- 9\times 2^{16} m^2 \pi^4 a^4\frac{1}{L^6}\sum_{\vec k_1,\vec k_2\ne0}
\frac{H(\vec k_1)^2 H(\vec k_2)^2}
{k_1^2\left[k_1^2+k_2^2 +(\vec k_1+\vec k_2)^2\right] k_2^2}
+ \cO(1/L) \,.
\label{eq:I2num}
\end{align}
Here the two terms have canceling logarithmic divergences.

\begin{comment}
The final quantity is given, in the notation of Ref.~\cite{\HSTH},
$S_I =\sum_{n=3}^\infty I_n$. The infinite sum can be recast as an integral
equation
\begin{multline}
S_I (\vec k, \vec p) =  
 \int_{\vec k_1,\vec k_2,\vec k_3} 
\cM_2(\vec k) G^\infty(\vec k,\vec k_1) 
\frac{\cM_2(\vec k_1)}{2\omega_{k_1}} G^\infty(\vec k_1,\vec k_2) 
\frac{\cM_2(\vec k_2)}{2\omega_{k_2}} G^\infty(\vec k_2,\vec k_3) 
\frac{\cM_2(\vec k_3)}{2\omega_{k_3}} G^\infty(\vec k_3,\vec p)  \cM_2(\vec p)
\\
-
 \int_{\vec k_1}  \cM_2(\vec k) G^\infty(\vec k,\vec k_1)\frac1{2\omega_{k_1}}  S_I(\vec k_1,\vec p)
\,,
\label{eq:SI}
\end{multline}
and the quantity we want is $S_I=S_I(0,0)$ evaluated with $E=3m$.
This equation can be discretized
into a matrix equation as for $\cD^{(u,u)}$, and the solution gives
\begin{equation}
\wt S_I = -9 L^3 
\left[
\frac1{1+ 2 \omega \cM_2^s \slashed{G}} 
\left(2 \omega \cM_2^s \slashed{G} \right)^4 
2\omega \cM_2^s\right]_{00}
+ \cO(1/L) \,.
\label{eq:SInum}
\end{equation}}
\end{comment}

The final quantity is 
$S_I =\sum_{n=3}^\infty I_n$, where $I_n$ is defined in Eq.~(124) of Ref.~\cite{\HSTH}.
 Given this definition, it is straightforward to evaluate the geometric sum to arrive at  
 \begin{equation}
S_I = 9 L^3 
\left[
\frac1{1+ 2 \omega \cM_2^s \slashed{G}} 
\left(2 \omega \cM_2^s \slashed{G} \right)^4 
2\omega \cM_2^s\right]_{00}
+ \cO(1/L) \,.
\label{eq:SInum}
\end{equation}

In the numerical evaluation of $F_3^\infty$, $\cL(\vec k)$, $\wt I_1$, $\wt I_2$ and $S_I$,
we can use the same group-theoretical simplifications as described in the numerical
solution of the quantization conditions. 
Thus the matrices involved have dimensions given
by the number of momentum shells. 
{We also note that $\wt I_1$
is the simplest of the quantities to calculate, 
since it involves only a column of $\slashed{G}$ rather than the full matrix.}

\section{Evaluation of $\widetilde F^s$}
\label{app:Fs}

In this appendix we describe how we numerically evaluate the
two versions of $\wt F^s$ that we use. Since both differ from the more standard
choice of Refs.~\cite{Luscher:1991n1,kari:1995}, 
based of zeta-function regulation, we think it is useful
to present a short description.

We begin with $\wt F^s_{\rm HS}$.
We rewrite Eq.~(\ref{eq:Fsa}) as
\begin{align}
\wt F^s(\vec k) &= \wt F^s_r(\vec k) + \delta\wt F^s(\vec k)\,,
\\
\widetilde F_{\rm HS}^s(\vec k) &=\frac{1}{2\omega_k} 
\frac{H(\vec k)}{32 \pi^3 (E-\omega_k)} \frac{2\pi}{L}
\left[\sum_{\vec n_a} - \,\PV\!\int_{\vec n_a} \right]
\frac{H(\vec a) H(\vec b)}{x^2 - r^2}
\,,
\end{align}
where $\delta \wt F^s$ is exponentially suppressed.
$\wt F_{\rm HS}^s$ is the form introduced in Eqs.~(43) and (44) of Ref.~\cite{\HSTH}, and contains 
a sum and integral over the vector of integers $\vec n_a$,
with $\vec a=(2\pi/L) \vec n_a$ and $\vec b=-\vec a - \vec k$,
while $x^2 \equiv q_{2,k}^{*\,2} L^2/(4\pi^2)$ is a quantity that can have either sign.
Finally, $r$ is magnitude of a vector whose parts parallel and perpendicular to
$-\vec k$ are
\begin{equation}
r_\parallel = \frac{n_{a\parallel}- |\vec n_k|/2}{\gamma}\,,
\ \
r_\perp = n_{a\perp}\,,
\end{equation}
with $\gamma=(E-\omega_k)/E_{2,k}^*$.

The reason for this rewriting is that it is now easier to numerically implement
 the PV prescription, following a method introduced in Appendix A of Ref.~\cite{\HSTH}.
Using $d^3n_a = \gamma d^3r$, the integral can be rewritten as
\begin{align}
\PV\!\int_{\vec n_a}
\frac{H(\vec a) H(\vec b)}{x^2 - r^2} &=
 \gamma
\PV\!\int_{\vec r}\frac{H(\vec a) H(\vec b)}{x^2 - r^2} 
\\
&=\gamma\left[- \int_{\vec r} \frac{H(\vec a) H(\vec b)}{r^2}
+ x^2 \int_{\vec r} \frac{H(\vec a) H(\vec b)-1}{r^2(x^2 - r^2)}
+ x^2 \,\PV \!\int_{\vec r} \frac1{r^2 (x^2-r^2)}\, \right].
\label{eq:HStrick}
\end{align}
The pole prescription is needed only for
the third integral on the right-hand side (rhs), and this integral vanishes identically
for all real $x^2$ (of either sign). The first integral on the rhs is IR finite, while in
the second the pole at $x^2=r^2$ is cancelled by the difference in the numerator.
Thus both integrals are straightforward to evaluate numerically, which we do
by breaking them up into regions where $H(\vec a)H(\vec b)$ vanishes, is nontrivial,
or equals unity.

For the calculations presented in the main text, we have dropped 
the quantity $\delta \wt F^s$, since it is exponentially suppressed.
This is theoretically consistent, since the quantization
condition is only accurate up to exponentially suppressed corrections in the first place.

We now turn to $\wt F^s_{\rm KSS}$. Here we again drop $\delta \wt F^s$, and
then use the exponential regulator of Ref.~\cite{Kim:2005gf} to define the sum and
integral
\begin{align}
\widetilde F_{\rm KSS}^s(\vec k) &= \lim_{\alpha\to 0} \frac{1}{2\omega_k} 
\frac{H(\vec k)}{32 \pi^3 (E-\omega_k)} \frac{2\pi}{L}
\left[\sum_{\vec n_a} - \,
\PV\!\int_{\vec n_a} \right]\frac{e^{\alpha(x^2-r^2)}}{x^2 - r^2}
\,.
\end{align}
Unlike for $\wt F^s_{\rm HS}$, where the $H$-functions cut off the sums and integrals
at finite values of $\vec n$, here both extend to infinite $|\vec n|$, albeit in a convergent fashion.
We cut off the sum when the contributions of higher terms drop below our desired precision
(roughly 1 part in $10^{11}$).
The integral, however, can be evaluated analytically, using the same method
as in Eq.~(\ref{eq:HStrick}) to take care of the principal value prescription:
\begin{align}
\PV\!\int_{\vec n_a} \frac{e^{\alpha(x^2-r^2)}}{x^2 - r^2}
&= \gamma\left[
-\int_{\vec r} \frac{e^{\alpha(x^2-r^2)}}{r^2}
+x^2 \int_{\vec r} \frac{e^{\alpha(x^2-r^2)}-1}{r^2(x^2 - r^2)}
+ x^2 \gamma\,\PV \!\int_{\vec r} \frac1{r^2 (x^2-r^2)}
\right]
\\
&=4 \pi \gamma\left[
-\sqrt{\frac{\pi}{4\alpha}} e^{-\alpha x^2}
+ \frac{\pi x}{2} {\rm erfi}\left(\sqrt{\alpha x^2}\right)
\right]\,,
\end{align}
where ${\rm erfi}(z)$ is the imaginary error function, defined by $d\,{\rm erfi}(z)/dz = 2 e^{z^2}/\sqrt{\pi}$ and ${\rm erfi}(0)=0$.
The final issue is how to take the limit $\alpha\to 0$.
As shown in Ref.~\cite{Kim:2005gf}, the $\alpha$ dependence comes in the
form $e^{-c L^2/\alpha}$, where $c=\cO(1)$.
Thus for sufficiently small $\alpha$ the corrections become numerically negligible.
We find that, for the range of values of $L$ that we use, taking 
$\alpha=0.5-1$ suffices for the desired numerical accuracy.

\bibliographystyle{apsrev4-1} %%% physical review
\bibliography{ref} %%% ref.bib file
	
\end{document}